\documentclass[12pt]{article}
\usepackage{amsmath}
\usepackage{graphicx}
\usepackage{enumerate}
\usepackage{natbib}
\usepackage{tikz}
\usepackage{url} % not crucial - just used below for the URL

\newcommand{\blind}{1}
\usepackage{xr}
\usepackage{amsmath}

\usepackage{xcolor}

\usepackage{lineno} 
\usepackage{xr-hyper}
\usepackage[subscriptcorrection]{newtxmath}
 \newcommand{\wh}{\widehat}
 \newcommand{\wt}{\widetilde}
\usepackage{threeparttable}
\usepackage{tabularx}
\usepackage{booktabs}
\usepackage{multirow}
\usepackage{enumitem}
\usepackage{bm}

\usepackage{amsthm}

\usepackage[noend,ruled,linesnumbered]{algorithm2e}
\usepackage{pgfplots}
\usepackage{tikz}
\usepgfplotslibrary{fillbetween}
\usetikzlibrary{shapes,decorations,arrows,calc,arrows.meta,fit,positioning}
\tikzset{
    -Latex,auto,node distance =1 cm and 1 cm,semithick,
    state/.style ={ellipse, draw, minimum width = 0.7 cm},
    point/.style = {circle, draw, inner sep=0.04cm,fill,node contents={}},
    bidirected/.style={Latex-Latex,dashed},
    el/.style = {inner sep=2pt, align=left, sloped}
}

\makeatletter
\newcommand{\figcaption}{\def\@captype{figure}\caption}
\newcommand{\tabcaption}{\def\@captype{table}\caption}
\makeatother

\newcommand{\cov}{{\rm \mathbb{C}ov}}

\newcommand{\var}{\mbox{$\mathbb{V}$ar}}

 \newcommand{\bA}{{\mathbf A}}
 \newcommand{\bB}{{\mathbf B}}
 \newcommand{\bC}{{\mathbf C}}
 \newcommand{\bD}{{\mathbf D}}
 \newcommand{\bF}{{\mathbf F}}
 \newcommand{\bE}{{\mathbf E}}

 \newcommand{\bI}{{\mathbf I}}
 
 \newcommand{\bL}{{\mathbf L}}
 \newcommand{\bM}{{\mathbf M}}
 
 \newcommand{\bP}{{\mathbf P}}
 \newcommand{\bR}{{\mathbf R}}
 \newcommand{\bS}{{\mathbf S}}
 \newcommand{\bU}{{\mathbf U}}
 \newcommand{\bV}{{\mathbf V}}
 \newcommand{\bW}{{\mathbf W}}
 \newcommand{\bX}{{\mathbf X}}
 \newcommand{\bY}{{\mathbf Y}}
 
 \newcommand{\bZ}{{\mathbf Z}}
 \newcommand{\ba}{{\mathbf a}}

 \newcommand{\bg}{{\mathbf g}}

 \newcommand{\bh}{{\mathbf h}}

 \newcommand{\bx}{{\mathbf x}}
 \newcommand{\by}{{\mathbf y}}
 \newcommand{\bz}{{\mathbf z}}
 
 \newcommand{\bbeta}  {\boldsymbol{\beta}}

 \newcommand{\blambda}{\boldsymbol{\lambda}}
 
 \newcommand{\bOmega}{\boldsymbol{\Omega}}

 \newcommand{\bSigma}{{\mathbf \Sigma}}

 \newcommand{\btheta} {\boldsymbol{\theta}}
 
 \newcommand{\bmu} {\boldsymbol{{\mathbf \mu}}}

 \newcommand{\beq}{\begin{eqnarray*}}
 	\newcommand{\eeq}{\end{eqnarray*}}
 \newcommand{\beqn}{\begin{eqnarray}}
 	\newcommand{\eeqn}{\end{eqnarray}}

\DeclareFontFamily{U}{mathx}{\hyphenchar\font45}
\DeclareFontShape{U}{mathx}{m}{n}{
      <5> <6> <7> <8> <9> <10> <10.95> <12> <14.4> <17.28> <20.74> <24.88>
      mathx10
}{}
\DeclareSymbolFont{mathx}{U}{mathx}{m}{n}
\DeclareMathAccent{\widecheck}{0}{mathx}{"71}

 \newtheorem{theorem}{Theorem}[section]
 \newtheorem{lemma}{Lemma}[section]
 \newtheorem{corollary}[theorem]{Corollary}

 \newtheorem{assumption}{Assumption}[section]
  
 \newtheorem{proposition}{Proposition}[section]

\newcommand{\diag}{\operatorname{diag}}

\pgfmathdeclarefunction{gauss}{2}{\pgfmathparse{1/(sqrt(2*pi)*#2)*exp(-((x-#1)^2)/(2*#2^2))}%
}

\pgfmathdeclarefunction{gauss3}{3}{%
        \pgfmathparse{1/(#3*sqrt(2*pi))*exp(-((#1-#2)^2)/(2*#3^2))}%
    }
    
\pgfmathdeclarefunction{chisqtrans}{1}{%
        \pgfmathparse{(x>(1-#1))*exp(-(x-(1-#1))/(2*(#1/3)))*sqrt(x-(1-#1))/sqrt(2*pi*(#1/3)^3)}%
}

\usepackage{setspace}
\usepackage{etoolbox}
\usepackage{subcaption} % loads caption automatically

\begin{document}

\if1\blind
{
  \title{\bf On the Consistency of Spectral Clustering under Generalized Gaussian Mixture Models}
  \author{Lyuou Zhang\\
    School of Statistics and Data Science\\
    Shanghai University of Finance and Economics\\
    Boxiang Wang\\
    Department of Statistics and Actuarial Science\\
    The University of Iowa\\
    Wen Zhou\\
    Department of Biostatistics, School of Global Public Health\\
    New York University\\
    Hui Zou\\
    Department of Applied Mathematics and Statistics\\ Johns Hopkins University}
  \maketitle
} \fi

\if0\blind
{
  \bigskip
  \bigskip
  \bigskip
  \begin{center}
    {\LARGE\bf On the Consistency of Spectral Clustering under Generalized Gaussian Mixture Models}
\end{center}
  \medskip
} \fi

\begin{abstract}
    Spectral clustering is a widely used clustering method with strong empirical performance and favorable computational scalability. Existing theory has established its consistency and risk guarantees under Gaussian and related mixture models, primarily in settings with relatively simple latent structure. In this paper, we study spectral clustering under a semiparametric Gaussian mixture model, where the observed data become Gaussian-mixture distributed only after unknown monotone coordinate-wise transformations. We derive non-asymptotic bounds for its cluster recovery error. As a byproduct of the theory, we show that spectral clustering provides a theoretically justified initialization for a broad class of high-dimensional mixture-based learning procedures.
\end{abstract}

\noindent%

{\it Keywords:} Consistency, g-mixture models,  Spectral clustering
\vfill

\def\spacingset#1{\renewcommand{\baselinestretch}%
{#1}\small\normalsize} \spacingset{1}
\spacingset{1.8} % 26 lines of text per page

 \section{Introduction}
 \label{sec:intro}

Spectral clustering is a leading method for unsupervised learning. It constructs a low-dimensional representation from the leading eigenspace of a similarity, Gram, or graph Laplacian matrix and then applies $K$-means to the resulting embedding \citep{NgJordanWeiss2001}. Because it does not require the specification of a parametric likelihood, spectral clustering can accommodate complex cluster geometry and has been used successfully for heterogeneity analysis in precision medicine and marketing \citep{fan2014challenges} and community detection in network analysis \citep{ke2024optimal}. Lately, it  has also been considered as an effective initialization for more elaborate likelihood-based or EM-type procedures. Statistical guarantees for spectral clustering are therefore important both when it is used as a final clustering method and when it serves as the first stage of a broader learning procedure.

As the algorithm itself does not posit a probabilistic model, the statistical analysis on spectral clustering typically proceeds under an assumed data-generating mechanism, with the Gaussian mixture model serving as a canonical benchmark. Early work established convergence of empirical spectral partitions to their population counterparts \citep{von2008consistency}, while subsequent advances in random matrix theory and entrywise spectral perturbation analysis yielded increasingly sharp bounds for eigenvector estimation and mis-clustering risk \citep{abbe2020entrywise,ndaoud2022sharp,amini2021concentration}. More recent studies extended such guarantees to sub-Gaussian or sub-exponential mixture models and established convergence rates and minimax lower bounds for the mis-clustering risk \citep{yan2016robustness,abbe2022,zhang2024leave}. These extensions, however, remain to impose strong restrictions on marginal shapes and tails and do not accommodate nonlinear coordinate-wise distortions. Such distortions are common in biomedical and genomic data, where observed features can be highly skewed, heavy-tailed, or nonlinearly transformed. Moreover, because these transformations alter the Euclidean geometry from which the spectral embedding is constructed, existing guarantees under Gaussian mixtures do not transfer directly.

An additional challenge is how to scale the spectral embedding, as this choice can materially affect both the accuracy and robustness of spectral clustering \citep{Jin2015,jin2022improvements}. Standard spectral clustering uses a $K$-dimensional embedding when there are $K$ clusters. However, the rank $r$ of the population signal matrix can be strictly smaller than $K$; when $r<K$, the remaining empirical spectral directions have no population signal counterpart and may be dominated by noise. Normalization offers a potential way to reduce the resulting instability, but existing theory does not provide a general characterization of when or why it improves clustering. Pre-spectral methods, such as degree or row-sum normalization of a graph Laplacian, are primarily designed to address degree heterogeneity in networks and generally rely on a nonnegative similarity matrix \citep{NgJordanWeiss2001,RoheChatterjeeYu2011,QinRohe2013,LeiRinaldo2015,JosephYu2016,fan2025asymptotic}. Post-spectral methods, including eigenvector and spherical normalization, are more broadly applicable, but their theoretical benefits remain model-dependent and have been studied mainly in network and Gaussian-mixture settings \citep{Jin2015,jin2022improvements,ke2024entryWise,ke2024optimal,agterberg2025joint}. In particular, unknown monotone coordinate-wise transformations alter the geometry and eigenstructure on which the spectral embedding depends, so existing results do not determine whether normalization can still mitigate the mismatch between $K$ and $r$. Treating these transformations as nuisance components rather than targets of identification, we establish spectral-clustering guarantees under a semiparametric Gaussian mixture model and characterize clustering performance under a self-normalized embedding.

\subsection{The $g$-mixture framework and our contributions}

The preceding discussion motivates studying spectral clustering under a model that separates latent cluster structure from nonlinear distortions in the observed data. For this purpose, we adopt the semiparametric $g$-mixture model introduced by \citet{wang2021clustering}. 
The $g$-mixture model was originally introduced as the foundation for CESME, a model-based high-dimensional clustering procedure that has demonstrated strong empirical performance, including favorable comparisons with widely used pipelines such as Seurat in single-cell RNA sequencing applications \citep{hao2024dictionary}. Its combination of distributional flexibility and latent structure also makes it a natural framework for studying spectral clustering: the unknown coordinate-wise transformations accommodate skewness, heavy tails, and nonlinear marginal distortions in the observed data, while the latent Gaussian mixture provides a well-defined cluster structure for theoretical analysis. 

Formally, let $z_i^*\in\{1,\ldots,K\}$ denote the latent component label of observation $i$, with $\mathbb{P}(z_i^*=k)=\lambda_k^*$ and $\lambda_k^*\in(0,1)$ for $\sum_{k=1}^K\lambda_k^*=1$. Let $\bg^*=(g_1^*,\ldots,g_p^*)$ be a collection of unknown strictly increasing coordinate-wise transformations. Conditional on $z_i^*=k$, assume that
\begin{equation}
\bg^*(\bX_i)| z_i^*=k \sim \mathcal{N}(\bmu_k^*, \bSigma^*),
\label{eq:cesme}
\end{equation}
where $\bg^*(\bX_i)=(g_1^*(X_{i1}),\ldots,g_p^*(X_{ip}))^\top$. We name model \eqref{eq:cesme} the $g$-mixture model, where the letter "g" means generalized and also stands for these $g$ functions.
In the $g$-mixture model, the transformed observations follow a Gaussian mixture with component-specific means and a common covariance matrix. The standard Gaussian mixture model is recovered when each $g_\ell^*$ is the identity map, whereas nonlinear transformations allow the observed coordinates to exhibit skewness, heavy tails, and nonlinear marginal distortions. In \citet{wang2021clustering}, it is assumed that for some $\delta>0$, $\max_{1\leq \ell\leq p}\sum_{k=1}^K\lambda_k^*\int_{-\infty}^{\infty} |x|^{4+\delta}
\frac{1}{\sigma_\ell^*}\phi\left(\frac{g_\ell^*(x)-\mu_{k\ell}^*}{\sigma_\ell^*}\right)d g_\ell^*(x) \leq M_{4+\delta}<\infty$, 
where $\phi$ is the density of a standard normal distribution, $(\sigma_\ell^*)^2$ is the $\ell$th diagonal entry of $\bSigma^*$, and $d g_\ell^*(x)$ denotes the Lebesgue--Stieltjes measure induced by $g_\ell^*$. The bound $M_{4+\delta}$ is taken to be uniform over the coordinates. This moment condition in fact is equivalent to $\max_{1\leq\ell\leq p} \mathbb{E}|X_{i\ell}|^{4+\delta} \leq M_{4+\delta}$, so that the observed variables may remain substantially skewed or heavy-tailed, but their tails cannot be so extreme that the fourth-order moments required for examining the observed Gram matrix fail to exist. This condition provides the moment control needed to establish concentration of $\bX^\top\bX$ and its empirical eigenvalues and eigenvectors. Following \citet{wang2021clustering}, we also adopt the coordinate-wise scale normalization, so that 
$\sum_{k=1}^K\lambda_k^*\mu_{k\ell}^*=0$, and $(\sigma_\ell^*)^2+\sum_{k=1}^K\lambda_k^*(\mu_{k\ell}^*)^2=1$ for each $\ell=1,\ldots,p$. These constraints fix the location and scale of the latent representation, so that $g_\ell^*(X_{i\ell})$ has marginal mean zero and marginal variance one. They are used only to define the $g$-mixture representation and its theoretical complexity unambiguously. 
Together, these assumptions explain why the $g$-mixture model is particularly suitable for the present analysis. It allows substantial departures from Gaussianity on the observed scale while retaining sufficient moment structure to study spectral clustering applied directly to the observed data.

In this paper, we use the $g$-mixture as the data-generating framework and apply spectral clustering directly to the observed data, without estimating $\bg^*$ or fitting the mixture parameters. This formulation raises a distinct theoretical question because monotone transformations preserve the latent labels but generally alter the Euclidean and Gram-matrix geometry on which spectral clustering relies, so existing Gaussian-mixture guarantees do not apply immediately. We establish conditions under which the observed-data spectral embedding nevertheless retains sufficient cluster information for consistent recovery and quantify how the unknown transformations affect clustering accuracy. The resulting theory reveals a previously unexplored robustness of spectral clustering to nonlinear distributional distortions, and more importantly, provides the foundation for the self-normalized embedding developed below.

Our first principal contribution is a nonasymptotic analysis of the misclustering error of spectral clustering under the $g$-mixture model. Whereas existing theory has focused primarily on Gaussian, sub-Gaussian, or sub-exponential mixtures \citep{yan2016robustness,abbe2022,zhang2024leave}, we analyze spectral clustering applied directly to observations related to a latent Gaussian mixture through unknown coordinate-wise monotone transformations. Using leave-one-out perturbation techniques \citep{fan2018formula}, we control the empirical eigenvalues and eigenvectors of the observed Gram matrix and derive explicit bounds for the resulting misclustering error. These bounds quantify how clustering accuracy depends on the transformation complexity $C_{\bg}$, ambient dimension, latent cluster separation, mixing proportions, and within-component covariance. The resulting theory reveals a surprising robustness: even when the observed distributions depart substantially from Gaussian or sub-Gaussian mixtures, spectral clustering remains effective and achieves consistent cluster recovery under certain conditions. We further extend the analysis to spectral clustering based on Lipschitz kernel matrices.

Our second contribution is to establish spectral clustering as a theoretically justified initialization method for a broad class of high-dimensional mixture-based learning procedures. Initialization is a fundamental issue in nonconvex clustering problems, particularly when the number of clusters exceeds two, since even relatively simple mixture models may admit poor local optima \citep{jin2016local}. We show that the estimators produced by Algorithm~\ref{alg:initial} are sufficiently accurate to satisfy the local initialization requirements needed for subsequent iterative refinement. This result provides theoretical support for spectral initialization in EM-type methods and more general alternating-maximization procedures. In particular, our guarantees apply to high-dimensional clustering methods such as CHIME \citep{cai2019chime} and CESME \citep{wang2021clustering}. More broadly, the result provides a theoretical explanation for the commonly observed effectiveness of spectral clustering as an initialization step for structured high-dimensional mixture-learning procedures.

\subsection{Notation}

For a vector $\ba$, denote by $\Vert \ba\Vert_1$ and $\Vert \ba\Vert_2$ the $\mathcal{L}_1$ and $\mathcal{L}_2$ norms, respectively. For vector $S\subset\{1,\dots,p\}$, denote by $|S|$ the cardinality of $S$. For a vector $ \ba\in\mathbb{R}^p$ and $S\in\{1,\dots,p\}$, denote by $\ba_S$ the vector in $\mathbb{R}^p$ that has the same coordinates as $\ba$ in $S$ and zero coordinates on the complement $S^c$ of $S$. For $s<p$, denote by $\Gamma(s)=\{\ba\in\mathbb{R}^p:2\Vert \ba_{S^c}\Vert_1\le 4\Vert \ba_{S}\Vert_1+3\sqrt{s}\Vert a\Vert_2, \mbox{ for } S\subset\{1,\dots,p\} \mbox{ and } |S|=s\}$. For vector $\ba$, denote $\Vert \ba\Vert_{2,s}=\sup_{\bx\in\Gamma(s),\Vert \bx\Vert_2=1}|\ba^{\top}\bx|$. For two vectors $\ba_1,\ba_2$, denote $\Theta(\ba_1,\ba_2)=\cos^{-1}(\ba_1^{\top}\ba_2/\{\Vert\ba_1\Vert_2\Vert\ba_2\Vert_2\})$ the angle between $\ba_1$ and $\ba_2$. For a $p\times p$ matrix $\bM$, denote by $\|\bM\|_{\mathbb{F}}$ the Frobenius norm, $\sigma_{\min}(\bM)$ the smallest singular value of $M$,  $\sigma_{\max}(\bM)$ the largest singular value of $\bM$, $\Vert \bM\Vert_{2}=\sup_{\Vert \bx\Vert_2=1}\Vert \bM \bx\Vert_2$, and $\Vert \bM\Vert_{2,s}=\sup_{\bx\in\Gamma(s),\Vert \bx\Vert_2=1}\Vert \bM \bx\Vert_2$. 
For $\bx=(x_1,\ldots,x_p)^{\top}$, write $\bg(\bx)=\{g_1(x_1),\ldots,g_p(x_p)\}^{\top}$.
For a deterministic sequence $a_{n,p}\in(0,1/2)$, define $\mathcal{D}_{n,p}=\{\bx\in\mathbb R^p:a_{n,p}\le\Phi(\{g_\ell^*(x_\ell)-\mu_{k\ell}^*\}/\sigma_\ell^*)\le1-a_{n,p}\text{ for each }k=1,\ldots,K\mbox{ and }\ell=1,\ldots,p\}$. The sequence $a_{n,p}$ is chosen so that $\mathcal{D}_{n,p}$ is nonempty; further requirements are specified in Section \ref{sec:main_proof}. For functions $\bg$ bounded on $\mathcal{D}_{n,p}$, define the restricted
supremum seminorms $\Vert\bg\Vert_{\infty}=\max_{1\le\ell\le p}
\sup_{\bx\in\mathcal{D}_{n,p}}|g_\ell(x_\ell)|$ and $\Vert\bg\Vert_{\infty,s}=\sup_{\ba\in\Gamma(s),\Vert\ba\Vert_2=1}\sup_{\bx\in\mathcal{D}_{n,p}}|\ba^{\top}\bg(\bx)|$. 
Denote $\blambda=(\lambda_1,\dots,\lambda_K)^{\top}\in\mathbb{S}^K:=\{\blambda\in\mathbb{R}^K:\sum_{k=1}^K\lambda_k=1\}$ and $\bmu=(\bmu_1^{\top},\dots,\bmu_K^{\top})^{\top}$. For $\bbeta=(\blambda,\bmu,\bSigma)\in\mathbb{S}^K\times\mathbb{R}^{pK+p^2}$, define 
$\Vert\bbeta\Vert_{2,s}^2=\Vert \blambda\Vert_2^2+\Vert\bmu\Vert_{2,s}^2+\Vert\bSigma\Vert_{2,s}^2$.

\section{Spectral clustering with self-normalized embedding}
  \label{sec:spectral} 

\subsection{Our method}

We first introduce the spectral clustering procedure considered in this paper. 
Under model \eqref{eq:cesme}, after the latent monotone transformation $\bg^*$, the covariance decomposes into a between-cluster component and a within-cluster component, $\text{cov}(\bg^*(X))=\bS+\bSigma^*$, where $\bS=\sum_{k=1}^K\lambda_k^*\bmu_k^*(\bmu_k^*)^{\top}$ has rank at most $K$. Since $\bSigma^*$ is homogeneous across clusters, all the clustering structure of model \eqref{eq:cesme} is entirely determined by the cluster means $\bmu_k$'s. Thus, all information for distinguishing clusters is encoded in $\text{cov}(\bg^*(X))$ is carried out by $\bS$. Since the column space of $\bS$ is precisely the subspace spanned by the population cluster means, recovering the eigenspace of $\bS$ is equivalent to estimating the population discriminative subspace. This naturally motivates estimating this subspace through principal component analysis. Because the latent transformation $\bg^*$is unknown, the population quantity $\bS$ cannot be directly estimated. We therefore construct a spectral embedding from the observed data matrix $\bX$. Although the latent Gaussian representation is unobservable, the low-rank structure responsible for clustering remains encoded in the observed Gram matrix. Also, since the nonzero eigenvalues of $\bX\bX^{\top}$ and $\bX^{\top}\bX$ coincide, and the right singular vectors provide the spectral embedding of the observations, we work with the Gram matrix $\bA=\bX^{\top}\bX$.

Additionally, the cluster means $\bmu_k^*$'s are not necessarily orthogonal to each other. 
In fact, since we assume $\sum_{k=1}^K\lambda_k^*\bmu_k^*=0$, the column space spanned by $\bmu_1^*,\dots,\bmu_K^*$ has rank at most $K-1$.
Also, $\bmu_1^*,\dots,\bmu_K^*$ may have different $\ell_2$ norms, so the informative spectral directions of $\bA=\bX^{\top}\bX$ induced by $\bmu_k^*$'s may have heterogeneous strengths, especially under the distortions induced by unknown $\bg^*$. Thus, a classical spectral clustering can be unstable and overly sensitive to weaker directions that are relevant to noise. 
Thus, in Algorithm \ref{alg:initial}, we employ self-normalization of the Gram matrix $\bA$. In particular, with the rank-$K$ eigendecomposition of $\bA$, denoted by $\wh \bV \wh \bD^2 \wh \bV^\top$, we normalize $\wh \bD$ as $\wt \bD=(\wh \bD^2+\alpha  \wh d_{\max}^2\bI)^{-1/2}\wh \bD$, where $\wh d_{\max}^2$ is the largest eigenvalue of $\bA$.
The regularization through $\alpha >0$ addresses this issue by upweighting directions relevant to signal and downweighting directions relevant to noise, thereby improving the stability of the spectral representation and making the resulting clustering more robust.  
The proposed algorithm is summarized in Algorithm~\ref{alg:initial}.

\allowdisplaybreaks
\begin{algorithm}[h]
\caption{Spectral clustering with self-normalized embedding}
\label{alg:initial}
\KwIn{Observations $\bX=(\bX_1,\dots,\bX_n)\in\mathbb{R}^{p\times n}$, prespecified number of clusters $K$, and normalization parameter $\alpha$.}
\KwOut{Cluster assignments $\wh \bz=(\wh z_1,\ldots, \wh z_n)^\top$.}

Construct the Gram matrix $\bA=\bX^{\top}\bX$.

Compute the rank-$K$ eigendecomposition of $\bA$, denoted by $\wh \bV \wh \bD^2 \wh \bV^\top$, 
where $\wh \bV\in\mathbb{R}^{n\times K}$ contains the leading $K$ eigenvectors and $\wh \bD^2$ is the diagonal matrix of the corresponding eigenvalues. 

Define normalized $\wh \bD$ as $\wt \bD=(\wh \bD^2+\alpha  \wh d_{\max}^2\bI)^{-1/2}\wh \bD$, 
where $\wh d_{\max}^2$ is the largest eigenvalue of $\bA$. 

Apply $K$-means to the rows of $\wh \bV\wt \bD$ (equivalently, the columns of $\wt \bD\wh \bV^\top$) to obtain the cluster assignments $\wh \bz$. 

\end{algorithm}

Algorithm~\ref{alg:initial} differs from the standard spectral clustering in an important way. In particular, the standard spectral methods typically normalize or regularize the similarity matrix at the graph level, for example through a normalized or regularized Laplacian to account for degree heterogeneity \citep{NgJordanWeiss2001,RoheChatterjeeYu2011,LeiRinaldo2015,QinRohe2013,JosephYu2016}. Other methods, such as SCORE, instead remove multiplicative nuisance effects through ratios of leading eigenvectors \citep{Jin2015}. Algorithm~\ref{alg:initial} is also different from recent generalized-Laplacian approaches, e.g., \cite{fan2025asymptotic}, which imposes normalization on the similarity matrix prior to eigendecomposition through operators of the form $\bL^{-a}\bA \bL^{-a}$, where $\bA$ is an adjacency or similarity matrix, $\bL$ is the diagonal matrix of row sums of $\bA$, and $a$ is a positive constant. By contrast, Algorithm~\ref{alg:initial} performs a self-normalization directly on the empirical spectrum by replacing the raw eigenvalue matrix $\wh \bD$ with $\wt \bD$. This adaptively shrinks weak empirical spectral directions induced by noise while preserving the dominant ones induced by signal, yielding a stabilized embedding for subsequent clustering. 
The resulting normalization is therefore neither a degree correction nor a row normalization, but a data-dependent reweighting of the empirical spectral coordinates.

 \subsection{Cluster recovery}
 \label{sec:upperbound}
   
We now study the performance of Algorithm \ref{alg:initial} for recovering the cluster assignments under the latent semiparametric mixture model in \eqref{eq:cesme}. Recall that $z_1^*,\dots,z_n^*$ are interpreted as the true clustering labels in our data-generating process in Section \ref{sec:intro}. For the estimated clustering labels $\wh \bz$, their recovery accuracy is measured by: 
$\mathcal{L}(\wh \bz)=\min_{\Pi\in\mathcal{S}_K}n^{-1}\sum_{i=1}^n 1\{\Pi(\wh z_i)\ne z_i^*\},$
where $\mathcal{S}_K$ is the set of all label permutations of $\{1,\dots,K\}$ \citep{jin2016influential,zhang2024leave}.
We denote the signal strength of model \eqref{eq:cesme} by $R_{\min}=\min_{j\ne k}\Vert(\bSigma^*)^{-1/2}(\bmu_k^*-\bmu_j^*)\Vert_2$.
Then, we adapt the leave-one-out perturbation framework developed for sharp spectral clustering theory \citep{LofflerZhangZhou2021,zhang2024leave} to a setting in which the normalization itself is random and acts on the full empirical spectrum. This yields the following finite-sample characterization of the clustering error of Algorithm~\ref{alg:initial}.

\begin{theorem}\label{thm:kernel_spectral}
Suppose that $c_{\lambda}\le\lambda_k^*\le 1-c_{\lambda}$ for $k=1,\ldots,K$ and $c_{\Sigma}^{-1}\le\sigma_{min}(\bSigma^*)\le\sigma_{max}(\bSigma^*)\le c_{\Sigma}$. For each $\ell=1,\dots,p$, suppose that $h_{\ell}=(g_{\ell}^*)^{-1}$ is locally absolutely continuous and let $\xi_{\ell}$ denote the weak derivative of $h_{\ell}$ satisfying that $\int h_{\ell}(y)\allowbreak\psi'(y)dy=-\int \xi_{\ell}(y)\allowbreak\psi(y)dy$ for any test function $\psi\in \mathbb{C}_c^{\infty}(\mathbb{R})$, where $\mathbb{C}_c^{\infty}(\mathbb{R})$ denotes the class of infinitely differentiable functions with compact support on $\mathbb{R}$. Let $H_{\ell}(x)=(\sigma_{\ell}^*)^{-1}\int_{-\infty}^{\infty}h_{\ell}(u)\phi((u-x)/\sigma_{\ell}^*)du$, $J_{\ell}=[\min_k\mu_{k\ell}^*,\max_k\mu_{k\ell}^*]$, and $\mathcal{I}=\{\ell=1,\dots,p:\mu_{k\ell}^*\ne\mu_{k'\ell}^*\mbox{ for some }k\ne k'\}$.
Assume that
$$C_{\bg}^2=\frac{\max_{k,\ell}\int(\sigma_{\ell}^*)^{-1}\allowbreak\{\xi_{\ell}(y)\}^2\phi(\{y-\mu_{k\ell}^*\}/\sigma_{\ell}^*)dy}{[\inf_{\ell\in\mathcal{I}}\inf_{x\in J_{\ell}}H_{\ell}'(x)\}]^2}<\infty,$$ and $R_{\min}\ge8c_\lambda^{-1/2}KC_{\bg}c_{\Sigma}$. Fix $\alpha \in(0,1)$. Then, 
the clustering assignments $\widehat{\bz}$ in Algorithm \ref{alg:initial} satisfies that, with probability at least $1-O(n^{-1})$,
\begin{align}\label{eq:error_clus}
\mathcal{L}(\widehat{\bz})\le \frac{16C_{\bg}^2c_{\Sigma}^2(n+p)}{nR_{\min}^2},
\end{align}
and $
\mathbb{E}[\mathcal{L}(\widehat{\bz})]\le 16C_{\bg}^2c_{\Sigma}^2(n+p)(nR_{\min}^2)^{-1}+Cn^{-1}$,
where $C$ is a positive constant.
 \end{theorem}

Since the loss function $\mathcal{L}$ is computed over the complete training data $\bX$, increasing sample size $n$ leads to more information on recovering the clustering structure, as well as more difficulty on estimating clustering assignments for each subject of $\bX$. Thus, $\mathcal{L}$ converges to zero only when clustering separability $R_{\min}$ diverges. Similar phenomenon is also documented in Proposition 1 in \cite{zhang2024leave}.
Compared with Proposition 1 in \cite{zhang2024leave}, Theorem \ref{thm:kernel_spectral} extends the theory for spectral clustering to mixture of non-Gaussian distributions.
For completeness, we clarify the derivative appearing in Theorem \ref{thm:kernel_spectral}. Notice that for each $\ell$, $g_{\ell}^*$ is strictly increasing, so $h_{\ell}=(g_{\ell}^*)^{-1}$ is also strictly increasing, and is differentiable almost everywhere. In addition, since $h_{\ell}$ is locally absolutely continuous, the weak derivative $\xi_{\ell}$ coincides almost everywhere with the ordinary derivative of $h_{\ell}$, and $\xi_{\ell}$ may equivalently be interpreted as the almost-everywhere derivative of $h_{\ell}$. See, for example, \citet{brezis2011functional} for the relevant properties of one-dimensional Sobolev functions.

Although Algorithm~\ref{alg:initial} does not estimate $\bg^*$ explicitly, Theorem \ref{thm:kernel_spectral} depends on $\bg^*$ through $C_{\bg}$.
By Lemma~\ref{lem:g_approx} in the appendix, the numerator in $C_{\bg}^2$ controls the within-cluster covariance of the observed data, whereas its denominator controls the extent to which separation between the latent component means is preserved in the observed component means. Their ratio therefore accounts for both within-cluster variation and attenuation of between-cluster separation. More precisely, $C_{\bg}$ quantifies the effect of $\bg^*$ on signal-noise-ratio of model \eqref{eq:cesme} and thus the performance of spectral clustering under the model.
First, larger values of $C_{\bg}$ demand stronger latent cluster separation. Second, from \eqref{eq:error_clus}, $C_{\bg}^2$ acts as an additional complexity term alongside the covariance level $c_{\Sigma}$. 
Thus, $C_{\bg}$ reflects how strongly the transformations can locally distort the spectral geometry relevant for clustering. In this sense, the key issue is not global nonlinearity of the transformations, but rather the local behavior of their inverse maps near the component centers. 

Under conditions of Theorem \ref{thm:kernel_spectral}, since $R_{\min}\ge8c_\lambda^{-1/2}KC_{\bg}c_{\Sigma}$ and $c_{\lambda}\le 1/K$, we have a simplified upper bound $\mathbb{E}[\mathcal{L}(\widehat{\bz})]\le (1+p/n)/(4K^3)+C/n$. Thus, Theorem \ref{thm:kernel_spectral} also shows that, as a general-purpose procedure, spectral clustering in Algorithm \ref{alg:initial} is particularly effective in fixed or low dimensions, and it often remains competitive when the ambient dimension grows moderately. 
In contrast, for genuinely high-dimensional clustering problems, methods specifically designed to exploit sparsity can be more effective. We therefore regard the bound in \eqref{eq:error_clus} as most informative when $p$ is fixed or grows slowly with $n$, so that $p/n\to 0$. Under this regime and if the local derivatives of $\bg^*$ are uniformly bounded away from zero and infinity, then $C_{\bg}=O(1)$, and \eqref{eq:error_clus} retains the same $1/R_{\min}^2$-type scaling as in the Gaussian benchmark \citep[Proposition 3.1,][]{zhang2024leave}.

%%%%%%%%%%%%%%%%%%%%%%%%%%%%%%%%
 
\section{Further implications and extensions}
\subsection{Spectral clustering as initialization for mixture-based clustering}

An important consequence of the cluster-recovery guarantee in Theorem 2.1 is that the estimated labels can be used to construct accurate initial estimates of the latent mixture parameters and transformations. Combining Theorem 2.1 with concentration bounds for the resulting within-cluster empirical quantities yields a theoretically valid initialization for other mixture-based clustering algorithm. 
 
%----

Provable initialization is relatively well understood when $K=2$. In the two-component case, a location transformation can always be applied so that the cluster means are symmetric, and under this symmetry it is often sufficient that the initial mean estimator is not orthogonal to the true mean for algorithms, such as EM, to converge. For example, \citet{xu2016global} require that the initial mean $\widehat{\bmu}_1^{(0)}$ satisfy $(\bmu_1^*)^\top \widehat{\bmu}_1^{(0)}\neq 0$. This condition is easily met under random initialization \citep{meilua1998experimental}, since a randomly chosen initializer is orthogonal to $\bmu_1^*$ with probability zero. Moreover, \citet{wu2019randomly} show that EM with random initialization can achieve near-optimal performance when $O(n^{1/2})$ iterations, rather than a finite number of iterations, are allowed. On the other hand, \citet{hardt2015tight} proposed an algorithm based on the method of moments, known as the HP algorithm, which is used at initialization to achieve the required asymptotic rate without relying on many assumptions about the model. They studied both the methodological and theoretical properties of this approach.

The situation is considerably more challenging when $K>2$. \citet{jin2016local} show that, even for a one-dimensional $K$-component Gaussian mixture with $K\ge 3$, random initialization can lead with nontrivial probability to a poor local solution. The analysis of \citet{wu2019randomly} cannot be extended to $K\ge 3$, and the HP algorithm relies crucially on symmetry of the component means, making it difficult to generalize to mixtures with multiple components. More broadly, commonly used initialization methods for learning mixture models, such as $K$-means \citep{hartigan1979algorithm} and sparse clustering \citep{witten2010framework}, apply beyond the two-component setting but generally do not come with comparable provable guarantees. Motivated by this gap, we use Algorithm~\ref{alg:initial} to construct initial cluster assignments, initial cluster means, and an initial covariance estimator for subsequent EM-type procedures. For Gaussian mixture models, the initialization conditions in \citet{balakrishnan2017statistical} and \citet{hao18} are formulated in terms of consistent initial clustering assignments, and our results show that the Algorithm~\ref{alg:initial}-based spectral initialization provides such consistency even under the $g$-mixture model, which allows data to be substantially deviated from Gaussian.
CHIME \citep{cai2019chime} studies the high-dimensional Gaussian mixture model with a common covariance matrix, which is essentially model \eqref{eq:cesme} under the identity transformation. To obtain consistent clustering, CHIME requires the initial estimators $\wh\bmu_k^{(0)}$'s and $\wh\bSigma^{(0)}$ to satisfy
 \begin{align*}
 	\sum_{k=1}^K\Vert\wh\bmu_k^{(0)}-\bmu_k^*\Vert_{2,s}^2+\Vert\wh\bSigma^{(0)}-\bSigma^*\Vert_{2,s}^2\lesssim\frac{1}{s}.
 \end{align*}
Since CHIME assumes $s=o(n/\log(p))$, this condition forces the initializer to converge to the truth as $s\to\infty$. Such a requirement can be restrictive in practice, especially when $K>2$.

By contrast, as discussed in Section~\ref{sec:intro}, as a high-dimensional clustering algorithm based on the $g$-mixture model \eqref{eq:cesme} with unknown coordinate-wise transformation $\bg$, CESME \citep{wang2021clustering} achieves consistent clustering assuming
\begin{align}
\max\Big\{
\|\widehat{\bg}^{(0)}-\bg^*\|_{\infty,s}^2,\,
\|\widehat{\bbeta}^{(0)}-\bbeta^*\|_{2,s}^2
\Big\}
\le \frac{R_{\min}^2}{16},\label{eq:cseme_initial}
\end{align}
where $\bbeta^*=(\blambda^*,((\bmu_1^*)^\top,\ldots,(\bmu_K^*)^\top)^\top,\bSigma^*)$ collects the true mixture parameters in \eqref{eq:cesme}, $\bg^*$ are the true transformations. 
The initialization condition in \cite{wang2021clustering} only required the initializer to lie in a local neighborhood of the truth whose radius is determined by $R_{\min}$; it does not require consistency of the initializer itself. In particular, when $s=O(1)$, the initialization conditions in \citet{cai2019chime} and \citet{wang2021clustering} are of the same order. 
Specifically, the initialization of CESME is based on the spectral procedure in Algorithm~\ref{alg:initial} with linear kernel. 
Given the initial labels $\widehat{z}^{(0)}$ from Algorithm~\ref{alg:initial}, we can estimate the clustering weights $\wh \lambda_k^{(0)} = n^{-1}\sum_i 1(\wh z_{i}^{(0)}=k)$, cluster mean $\wt{\bmu}_k^{(0)}=\{\sum_i 1(\wh z_{i}^{(0)}=k)\}^{-1}\sum_i 1(\wh z_{i}^{(0)}=k)\bX_i$ for each $k$, and a pooled variance $(\wt \sigma_{\ell}^{(0)})^2=n^{-1}\sum_{k=1}^K\sum_{i:\wh z_{i}^{(0)}=k}(X_{i\ell}-\wt\mu_{k\ell}^{(0)})^2$ for each $\ell=1,\dots,p$. Then, we employ the initial transformation estimator 
$\widehat{g}_{\ell}^{(0)}(x) = \wt\sigma_{\ell}^{(0)}\sum\nolimits_{k=1}^K \widehat{\lambda}_k^{(0)}\allowbreak \Phi^{-1} \{ \widehat\psi_{\ell k}(x)\}$,
where $\widehat \psi_{\ell k}(x) = \{\sum_{i=1}^n 1(\widehat z_{i}^{(0)} = k)\}^{-1}\sum_{i=1}^n 1(X_{i\ell}\leq x, \widehat z_{i}^{(0)} = k)$. Lastly, we let $\wh{\bmu}_k^{(0)}=\{\sum_i 1(\wh z_{i}^{(0)}=k)\}^{-1}\sum_i 1(\wh z_{i}^{(0)}=k)\wh{\bg}^{(0)}(\bX_i)$ for each $k$, and $\wh {\bSigma}^{(0)}=n^{-1}\sum_{k=1}^K\sum_{i:\wh z_{i}^{(0)}=k} \{\wh{\bg}^{(0)}(\bX_i)-\wh\bmu_k^{(0)}\}
\{\wh{\bg}^{(0)}(\bX_i)-\wh\bmu_k^{(0)}\}^{\top}$ be the corresponding pooled covariance estimator based on the transformed observations. 
This initializer is shown to satisfy the initialization condition for CESME in \eqref{eq:cseme_initial}. More generally, the same argument applies to mixture-based procedures whose local convergence theory requires an initializer satisfying a neighborhood condition of the form \eqref{eq:cseme_initial}. When $\bg^*$ is the identity transformation, this includes Gaussian-mixture procedures such as CHIME \citep{cai2019chime}.

\begin{proposition}
\label{cor:initial_0}
Under the conditions of Theorem~\ref{thm:kernel_spectral}, using the region $\mathcal{D}_{n,p}$ specified in Section \ref{sec:main_proof} in the appendix, if $n$ is sufficiently large, then with probability at least $1-O(n^{-1})$,
\[
\max\Big\{
\|\widehat{\bg}^{(0)}-\bg^*\|_{\infty,s}^2,\,
\|\widehat{\bbeta}^{(0)}-\bbeta^*\|_{2,s}^2
\Big\}
\le \frac{R_{\min}^2}{16},
\] where $\widehat{\bbeta}^{(0)}=(\wh\blambda^{(0)}, ((\wh\bmu_1^{(0)})^{\top},\ldots, (\wh\bmu_K^{(0)})^{\top})^{\top}, \wh\bSigma^{(0)})$ are produced by Algorithm \ref{alg:initial}.
Consequently, the initial estimators $\wh{\bg}^{(0)}$, $\wh\bmu_1^{(0)},\ldots, \wh\bmu_K^{(0)}$, $\wh\blambda^{(0)}$, and $\wh\bSigma^{(0)}$ satisfy the initialization requirement for CESME; see Condition 3.3 of \citet{wang2021clustering}.
\end{proposition}

Thus, Proposition~\ref{cor:initial_0} gives a direct theoretical guarantee for the initialization used by CESME, Algorithm~2 of \citet{wang2021clustering}.  
The same spectral initialization can also apply to several Gaussian-mixture models. 
Indeed, when $\bg^*$ is the identity map, model \eqref{eq:cesme} reduces to a Gaussian mixture model, so Theorem~\ref{thm:kernel_spectral} also yields initialization guarantees for CHIME using Algorithm \ref{alg:initial} if $s=O(1)$, and for other procedures that require only control of the initial cluster means \citep{yi2015regularized,zhao2020statistical,wang2023statistical}.

In particular, the model considered by \citet{wang2023statistical} can be viewed as \eqref{eq:cesme} with identity transformation and they imposes the initialization condition $\max_k\Vert\wh\bmu_k^{(0)}-\bmu_k^*\Vert_{2}\le C R_{\min}$
for some constant $C>0$. 
The models studied by \citet{yi2015regularized} and \citet{zhao2020statistical} further assume identity covariance, so the signal strength is measured by $\min_{j\neq k}\|\bmu_k^*-\bmu_j^*\|_2$, and the initialization condition in \citet{zhao2020statistical} becomes
$\max_k\Vert\wh\bmu_k^{(0)}-\bmu_k^*\Vert_{2}\le \min_{j\ne k}\Vert\bmu_k^*-\bmu_j^*\Vert_2/2$.
The next proposition shows that the initializer from Algorithm~\ref{alg:initial} satisfies both initialization requirements.

  \begin{proposition}
 	\label{cor:initial_1}
 	Under the conditions of Theorem \ref{thm:kernel_spectral}, if $g_{\ell}^*(x)\equiv x$ for each $\ell=1,\dots,p$, $R_{\min}\ge8c_\lambda^{-1/2}KC_{\bg}c_{\Sigma}$ and $n$ is sufficiently large, then with probability at least $1-O(n^{-1})$, the initialization $\wh\bmu_k^{(0)}$'s produced by Algorithm \ref{alg:initial} satisfy that 
    \begin{enumerate}[label=(\roman*)]
        \item $\max_k\Vert\wh\bmu_k^{(0)}-\bmu_k^*\Vert_{2}\le R_{\min}/4$, and
        \item $\max_k\Vert\wh\bmu_k^{(0)}-\bmu_k^*\Vert_{2}\le \min_{j\ne k}\Vert\bmu_k^*-\bmu_j^*\Vert_2/8$.
    \end{enumerate}

  \end{proposition}

In Proposition \ref{cor:initial_1}, part (i) verifies the initialization condition in \cite{wang2023statistical}, while part (ii) guarantees the condition in \citet{zhao2020statistical}.
When $K=2$, the condition in \citet{yi2015regularized} is equivalent to that in \citet{zhao2020statistical} after centering and symmetrizing the initializer so that $\bmu_1^*=-\bmu_2^*$ and $\widehat{\bmu}_1^{(0)}=-\widehat{\bmu}_2^{(0)}$. Hence, Algorithm \ref{alg:initial} also provides legitimate initialization for \citet{yi2015regularized}.

Finally, \citet{xu2016global} and \citet{kwon2019global} study the two-component case under fixed dimension and require only that the initial mean estimator not be orthogonal to the true mean direction. That is, the initialization condition can be given as $(\bmu_1^*)^{\top}\wh\bmu_1^{(0)}\neq 0$, which rules out convergence of the likelihood to unfavorable local solutions. 
After centring and symmetrizing as above, Theorem~\ref{thm:kernel_spectral} yields the following guarantee.

  \begin{proposition}\label{cor:initial_2}
 Under the conditions of Theorem~\ref{thm:kernel_spectral}, if $g_{\ell}^*(x)\equiv x$ for each $\ell=1,\dots,p$, $R_{\min}\ge8c_\lambda^{-1/2}KC_{\bg}c_{\Sigma}$, $n$ is sufficiently large, and $K=2$, then the initialization $\wh\bmu_k^{(0)}$'s  produced by Algorithm \ref{alg:initial} satisfy that
$$
 		\mathbb{P}((\bmu_1^*)^{\top}\wh\bmu_1^{(0)}=0)\lesssim\frac{pC_{\bg}^2+nc_{\Sigma}}{nR_{\min}^2}.
$$
  \end{proposition} 
Proposition~\ref{cor:initial_2} implies that for fixed $p$, as $n\rightarrow\infty$, with high probability, the initial mean estimator $\widehat{\bmu}_1^{(0)}$ is not orthogonal to the true mean $\bmu_1^*$, so it satisfies the initialization conditions in \citet{xu2016global} and \citet{kwon2019global}, with high probability. We summarize the aforementioned initialization conditions and their theoretical justifications through Algorithm \ref{alg:initial} in Table \ref{tab:ini_cond}.

\begin{table}[!h]
\centering
\setlength{\tabcolsep}{4pt}
\begin{threeparttable}
\begin{tabularx}{\textwidth}{p{0.25\textwidth} >{\hsize=1.5\hsize\raggedright\arraybackslash}X >{\hsize=0.475\hsize\raggedright\arraybackslash}X}
\toprule
Reference & Initialization condition & Proof source \\
\midrule
{\citet{wang2021clustering}}
& {$\max\{\Vert\wh{\bg}^{(0)}-\bg^*\Vert_{\infty,s}^2,\ \Vert\wh\bbeta^{(0)}-\bbeta^*\Vert_{2,s}^2\}\le R_{\min}^2/16$}
& Pro. \ref{cor:initial_0} \\
{\citet{cai2019chime} }
& { $\sum_{k=1}^K\!\Vert\wh\bmu_k^{(0)}-\bmu_k^*\Vert_{2,s}^2+\Vert\wh\bSigma^{(0)}-\bSigma^*\Vert_{2,s}^2\!\lesssim\!1$ ($s=O(1)$)}
& Pro. \ref{cor:initial_0}  \\
\citet{yi2015regularized}
& $\Vert\wh\bmu_1^{(0)}-\bmu_1^*\Vert_{2}\le \Vert\bmu_1^*\Vert_{2}/4$ (two clusters)
& Pro.~\ref{cor:initial_1} (ii) \\
\citet{zhao2020statistical}
& $\max_k\Vert\wh\bmu_k^{(0)}-\bmu_k^*\Vert_{2}\le \min_{j\ne k}\Vert\bmu_k^*-\bmu_j^*\Vert_2/2$
& Pro.~\ref{cor:initial_1} (ii) \\

\citet{wang2023statistical} & $\max_k\Vert\wh\bmu_k^{(0)}-\bmu_k^*\Vert_{2}\le C R_{\min}$ & Pro.~\ref{cor:initial_1} (i) \\
\citet{hao18} 
&\multirow{2}{20em}{ $\mathcal{L}(\wh{\bz}^{(0)})=o_P(1)$ ($p/n\to 0$)}
& \multirow{2}{20em}{Thm.~\ref{thm:kernel_spectral}} \\
\citet{balakrishnan2017statistical}&  &  \\
\citet{xu2016global}
& $(\bmu_1^*)^{\top}\wh\bmu_1^{(0)}\neq 0$ (two clusters, $p/n\to 0$)
& Pro.~\ref{cor:initial_2} \\
\citet{kwon2019global} & $\cos(\Theta(\wh\bmu_1^{(0)},\bmu_1^*))\neq 1$ (two clusters, $p/n\to 0$) & Pro.~\ref{cor:initial_2} \\
\bottomrule
\end{tabularx}
\caption{Initialization conditions and proof sources for selected clustering methods (abridged). The column “Proof source” points to the theorem/proposition in this paper that verifies the stated initialization under the cited model.}
\label{tab:ini_cond}
\end{threeparttable}
\end{table}

\subsection{Extension to kernel spectral clustering}
\label{sec:ksc}

It is worth noticing that, data from $g$-mixture model in \eqref{eq:cesme} do not follow the Gaussian mixture when $\bg^*$ is not identity or linear transformation, but the transformed data $\bg^*(X)$ follow a Gaussian mixture model. That is, if we project the data into another space with some specific transformations, the projected data follows a Gaussian mixture model. In practice, projecting the data is computationally expensive, so we consider the kernel function relative to the transformation, which gives the proposed kernel spectral clustering, defined in Algorithm \ref{alg:initial}.
Notice that, if we specify the kernel function $\mathcal{K}^*$ corresponding to the transformation $\bg^*$, Algorithm \ref{alg:initial} with kernel $\mathcal{K}^*$ is equivalent to Algorithm \ref{alg:initial} with linear kernel on transformed data.

In practice, $\bg^*$ and $\mathcal{K}^*$ are always unknown.
Therefore, we consider using a general kernel function $\mathcal{K}$ to construct  kernel matrix $\bA^{\mathcal{K}}=\mathcal{K}(\bX,\bX)$ and apply Algorithm \ref{alg:initial}.
The use of a kernel is motivated by the latent semiparametric mixture model in expression~\eqref{eq:cesme}. 
Although the transformed data $\bg^*(\bX_i)$ follow a Gaussian mixture, the observable $\bX_i$ may deviate substantially from Gaussianity due to the unknown $\bg^*$.
In this case, a kernel matrix can better capture latent cluster separation than a purely linear embedding based on the raw observations. 

A kernel-based similarity matrix provides a flexible way to extract low-dimensional cluster structure from the observed data without explicitly estimating $\bg^*$. Hence, we consider a kernel function $\mathcal{K}:\mathbb{R}^p\times\mathbb{R}^p\to\mathbb{R}$ satisfying the following condition.
\begin{assumption}\label{cond:kernel}
 	The kernel function $\mathcal{K}$ is symmetric, positive semidefinite, and $\rho$-Lipschitz such that $
 		|\mathcal{K}(\bx_1,\bx_2)-\mathcal{K}(\by_1,\by_2)|\le \rho(\Vert \bx_1-\by_1\Vert_2+\Vert \bx_2-\by_2\Vert_2)$
 	for any $\bx_1,\bx_2,\by_1,\by_2\in\mathbb{R}^p$, where $\rho$ is a positive constant.
 \end{assumption}

Condition \ref{cond:kernel} imposes a mild smoothness assumption on $\mathcal{K}$ to be Lipschitz continuous with respect to the Euclidean norm in each argument. It is satisfied, for example, by distance-based kernels of the form $\mathcal{K}(\bx_1,\bx_2)=f(\Vert \bx_1-\bx_2\Vert_2)$, where $f:\mathbb{R}\rightarrow\mathbb{R}$ is $\rho$-Lipschitz. A canonical example is the Gaussian kernel, for which $f(t)\propto \exp\{-t^2/(2\tau^2)\}$ with $\tau\asymp \rho^{-1}$. Given observations $\bX=(\bX_1,\dots,\bX_n)$, we form the kernel matrix
$
\bA^{\mathcal{K}}=\mathcal{K}(\bX,\bX)=\big(\mathcal{K}(\bX_i,\bX_j)\big)_{i,j=1}^n:=\big(a_{ij}\big)_{i,j=1}^n.
$
This matrix can be used to construct the spectral clustering algorithm in Algorithm~\ref{alg:initial}. 
Also, we denote $B_{kk'}=\mathbb{E}[\mathcal{K}(\bX_i,\bX_{i'})| z_i^*=k,z_{i'}^*=k']$ for each $k,k'=1,\dots,K$ and $\bB=\{B_{kk'}\}_{k,k'=1}^K$.
Then, we have the following corollary on Algorithm~\ref{alg:initial} with general kernel.

\begin{corollary}\label{cor:kernel_spectral}
Under conditions of Theorem \ref{thm:kernel_spectral}, and assuming that kernel function $\mathcal{K}$ satisfies Condition \ref{cond:kernel} and $\sigma_{\min}(\bB)/\sigma_{\max}(\bB)>c_{\bB}$, where $c_{\bB}$ is a positive constant, the clustering assignments $\widehat{\bz}$ in Algorithm \ref{alg:initial} with $\bA^{\mathcal{K}}=\mathcal{K}(\bX,\bX)$ satisfies that, with probability at least $1-O(n^{-1})$,
\begin{align*}
\mathcal{L}(\widehat{\bz})\le \frac{16C_{\bg}^2c_{\Sigma}^2(n+p)}{nR_{\min}^2}.
\end{align*}
\end{corollary}

By Corollary \ref{cor:kernel_spectral}, with properly chosen kernel function $\mathcal{K}$, Algorithm \ref{alg:initial} with general kernel attains similar clustering performance as that with linear kernel. This allows Algorithm \ref{alg:initial} to be extended to a broader range of kernel spectral clustering. 
In particular, if $\mathcal{K}$ is the Gaussian kernel with bandwidth $\tau$, we have
\begin{align*}
c_{\bB}\ge\frac{1-(K-1)\exp\{-R_{\min}^2/(2\tau^2c_{\Sigma}+2)\}}{1+(K-1)\exp\{-R_{\min}^2/(2\tau^2c_{\Sigma}+2)\}},
\end{align*}
so that $c_{\bB}>0$ if $\tau^2c_{\Sigma}/R_{\min}^2\in[c_1,c_2]$ for some positive constants $c_1$ and $c_2$.
Thus, the bandwidth $\tau$ of Gaussian kernel function $\mathcal{K}$ cannot be arbitrarily chosen and is required to have the same order as $R_{\min}$. 
Also, for the Gaussian kernel, the Lipschitz constant $\rho$ is of order $\tau^{-1}$, so an excessively small bandwidth increases the perturbation sensitivity of the kernel matrix. Conversely, as $\tau$ becomes too large, the between-component kernel similarities become nearly indistinguishable, causing $\bB$ to approach a rank-one matrix and $\sigma_{\min}(B)/\sigma_{\max}(B)$ to deteriorate. Thus, the bandwidth must balance kernel smoothness against preservation of population cluster separation.

%%%%%%%%%%%%%%%%%%%%%%%%%%

\section{Numerical illustrations}

\subsection{Simulation studies}

We generated data from the $g$-mixture model in \eqref{eq:cesme} with dimension $p=20$ and sample size $n=100,200,\dots,900$. We consider a four-component mixture with equal mixing proportions, that is, $K=4$ and $\lambda_k=0.25$ for $k=1,\dots,4$. Two covariance structures are examined. The first is an inverse AR(1) model, $\bSigma=\bOmega^{-1}$, where $\Omega_{ij}=0.8^{|i-j|}$. The second is an asymptotic quadratic decay (AQD) model, in which the $(i,j)$th entry of $\bSigma$ is given by $0.7/(0.7 + |i-j|)^2$.
The cluster means $\bmu_k$ are specified by
$
\mu_{1\ell}=3\sqrt{10}$ for $\ell=1,\dots,p/K$, $\mu_{k\ell}=3/\sqrt{10}$ for $k=2,3,4$ and $\ell=(k-1)p/K+1,\dots,kp/K$, and $\mu_{k\ell}=0$ otherwise. Then, we generate observations $X_i\in\mathbb{R}^p$ for $i=1,\dots,n$ independently from the Gaussian mixture
$
\sum_{k=1}^K \lambda_k \mathcal{N}(\bmu_k,\bSigma).
$

To generate observations $\bX_i\in\mathbb{R}^p$, $i=1,\dots,n$, we first sample latent variables $\bZ_i$ independently from the Gaussian mixture
$
\sum_{k=1}^K \lambda_k \mathcal{N}(\bmu_k,\bSigma)
$,
and then apply a coordinate-wise inverse transformation to obtain $\bX_i$. Specifically, for each coordinate $\ell=1,\dots,p$, we set $X_{i\ell}=g^{-1}(Z_{i\ell})$. We consider three transformation settings: (1) arcsinh: $g^{-1}(x)=\sinh(x)$; (2) polynomial: $g^{-1}(x)=x+2x^3$; and (3) identity: $g^{-1}(x)=x$.
All datasets are centered before conducting clustering analysis.

We applied Algorithm \ref{alg:initial} with the Gram matrix $\bA=\bX^{\top}\bX$, where $\alpha$ ranges from $0$ to $0.1$. For competing methods, we consider spectral clustering on scaled eigenvectors \citep{zhang2024leave}, spectral clustering on unscaled eigenvectors with post-PCA spherical normalization \citep{NgJordanWeiss2001}, spectral clustering on scaled eigenvectors with post-PCA spherical normalization \citep{agterberg2025joint}, spectral clustering on scaled eigenvectors with post-PCA eigenvector normalization \citep{Jin2015}, and SCORE+ \citep{jin2022improvements}. All methods are conducted with the true number of clusters $K$. For each spectral clustering algorithm, the $K$-means algorithm is run with $10$ starting points and the result with smallest within-cluster variance is reported.  Since pre-PCA normalization is only applicable on kernel matrix constructed with non-negative kernel function, the comparison of Algorithm \ref{alg:initial} against spectral clustering with pre-PCA normalization is omitted.
\begin{figure}[!h]
    \centering
    \includegraphics[height=15cm, keepaspectratio]{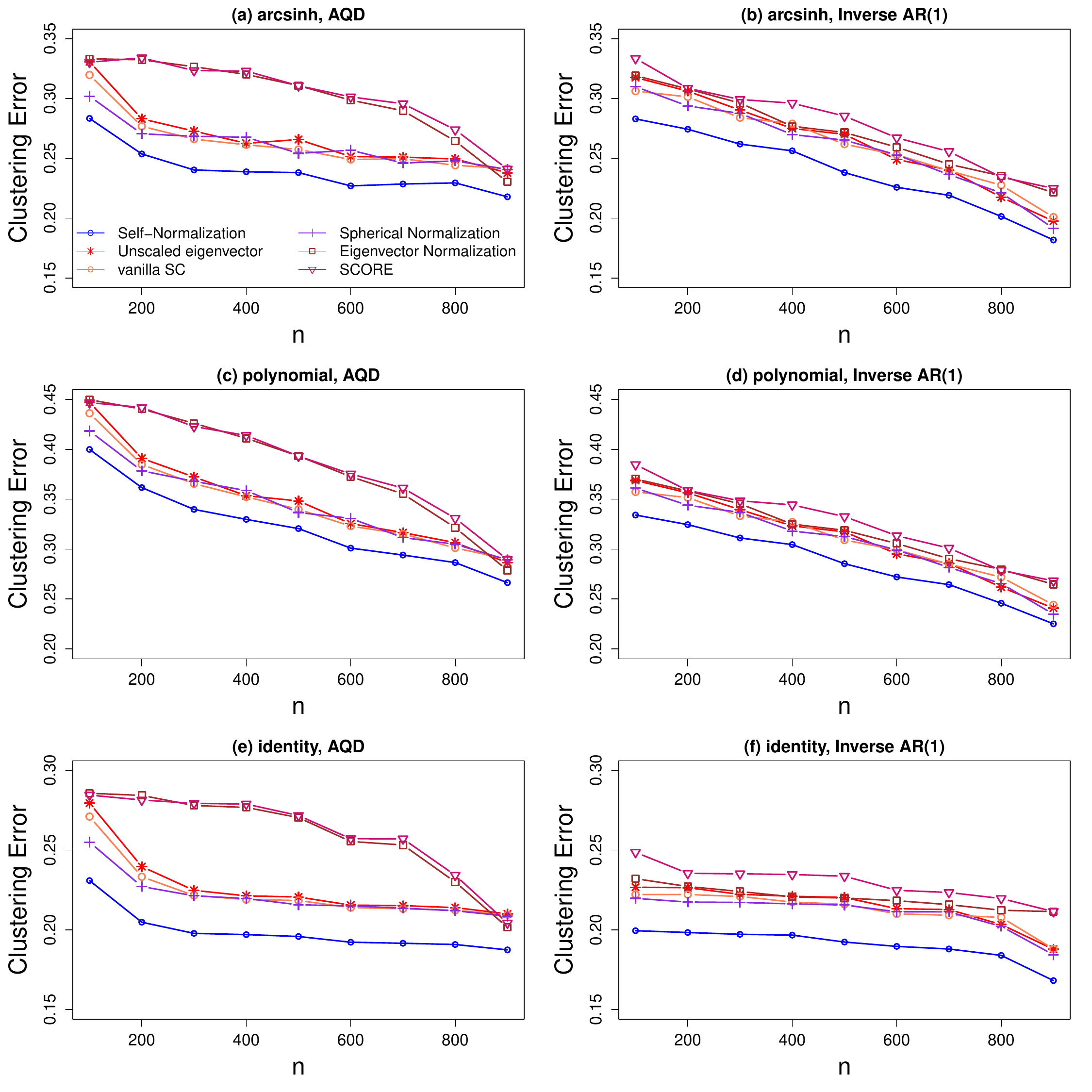}
    \caption{Comparison of clustering errors for Algorithm~\ref{alg:initial} against the competing methods. The dimension is fixed at $p=20$, and the sample size $n$ varies from $100$ to $900$. The transformation and covariance structure are indicated in the figure titles. Algorithm \ref{alg:initial} outperforms the competing methods with smallest clustering error.
    }
    \label{fig:clus_error}
\end{figure}

First, we consider Algorithm \ref{alg:initial} with $\alpha=0.05$ and competing methods on $\bA=\bX^{\top}\bX$. The results, based on $100$ independent replications, are reported in Figure~\ref{fig:clus_error}. As displayed in Figure~\ref{fig:clus_error}, Algorithm~\ref{alg:initial} outperforms other spectral clustering algorithms, which justifies the self normalization in Algorithm~\ref{alg:initial}. In addition, as the sample size $n$ increases, the clustering errors of all spectral clustering algorithms do not converge to $0$, but to a positive number depending on the covariance structure. This is because, although $n$ diverges, the separation among clusters, quantified by $R_{\min}$, stays constant. In fact, for fixed $R_{\min}$, clustering errors of spectral clustering algorithms are upper bounded by a non-vanishing positive constant, which is shown in Theorem \ref{thm:kernel_spectral}.

\begin{figure}[!h]
    \centering
    \includegraphics[height=15cm,
        keepaspectratio]{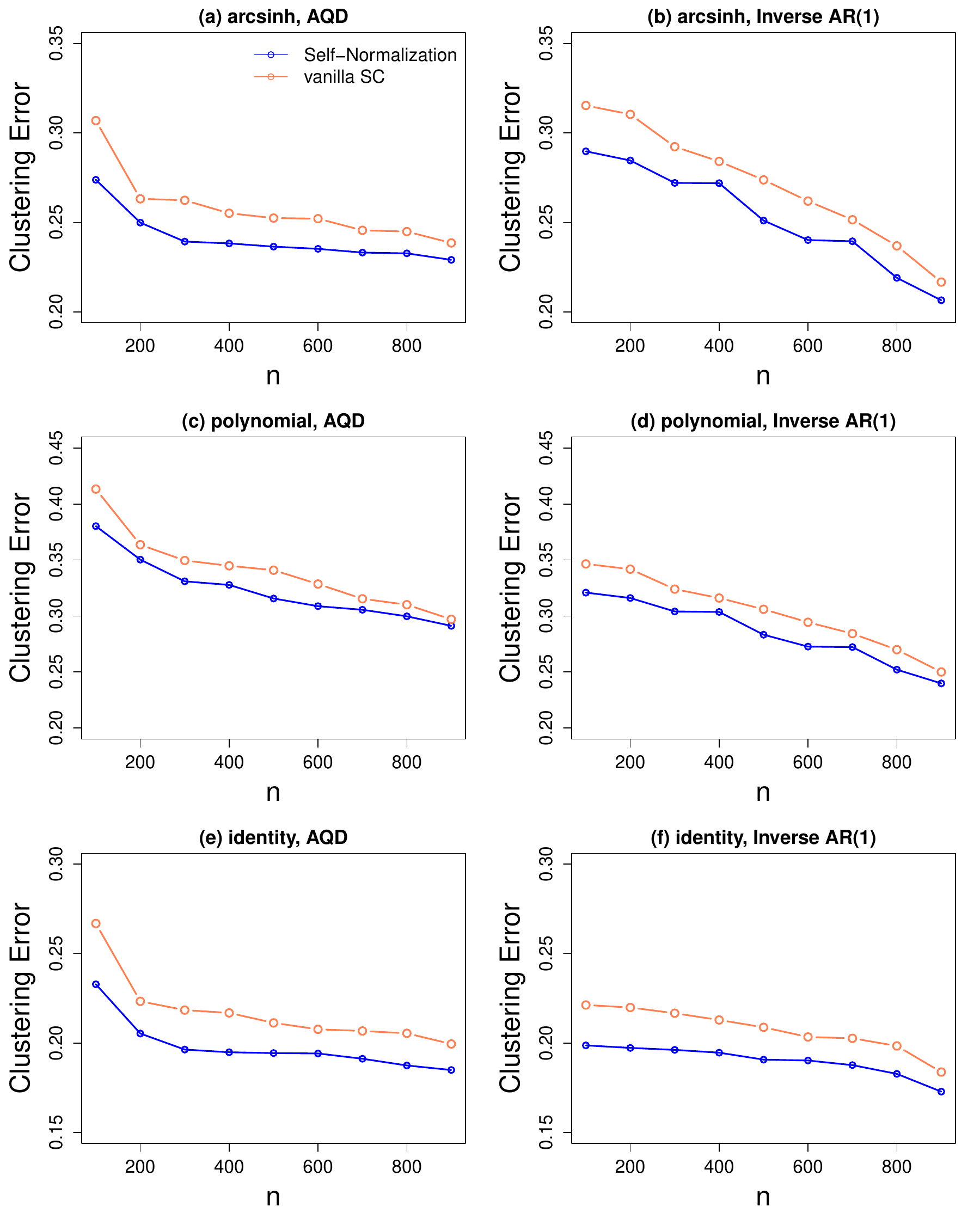}
    \caption{
    Comparison of clustering errors for Algorithm~\ref{alg:initial} against the competing methods. The dimension is fixed at $p=20$, and the sample size $n$ varies from $100$ to $900$. The transformation covariance structure are indicated in the figure titles. Algorithm \ref{alg:initial} outperforms vanilla spectral clustering with smaller clustering error.}
    \label{fig:clus_error_gaussian}
\end{figure}

Lastly, we compare Algorithm \ref{alg:initial} against vanilla spectral clustering on the kernel matrix $\bA=\mathcal{K}(\bX,\bX)$ with Gaussian kernel $\mathcal{K}(x_1,x_2)\propto\exp\{-\Vert x_1-x_2\Vert_2^2/2\}$. 
Since the theories on other normalization approaches are based on linear kernel, the comparison of Algorithm \ref{alg:initial} against these normalizations is omitted.
The results, based on $100$ independent replications, are reported in Figure~\ref{fig:clus_error_gaussian}. 
As displayed in Figure~\ref{fig:clus_error_gaussian}, Algorithm~\ref{alg:initial} outperforms vanilla spectral clustering, which shows that its good clustering performance is not affected by the choice of kernel functions.

 %%%%%%%%%%%%%%%%%%%%%%%%%%%%%%%%%%%%%%%%%%%

\subsection{Real data demonstration}

In this subsection, we compare our proposed algorithm with competing methods on real data.
We use the Wisconsin Diagnostic Breast Cancer (WDBC) dataset \citep{street1993nuclear}, Dermatology, Ecoli datasets from the UCI Machine Learning Repository \citep{asuncion2007uci}, and the Levine\_32dim single-cell mass-cytometry (Levine) dataset \citep{levine2015data} as real-data illustrations of the proposed method.
The WDBC dataset consists of quantitative cellular characteristics extracted from digitized breast-mass images with substantial heterogeneity and nonlinear dependence, making them a representative benchmark for evaluating clustering methods beyond ideal Gaussian settings.
Dermatological diagnosis relies on heterogeneous clinical attributes whose marginal distributions are often highly skewed, while Protein localization data are well known to contain nonlinear feature distributions arising from biological measurement processes. Thus, these datasets represent three different domains and evaluate robustness across diverse data structures.
The sample size $n$, number of features $p$, and number of clusters $K$ of each dataset are listed in Table \ref{tab:wdbc_clus}. All datasets are centralized before conducting clustering analysis.

We choose $\alpha=0.05$ for Algorithm \ref{alg:initial}. 
For competing methods, we still consider spectral clustering on scaled eigenvectors \citep{zhang2024leave}, spectral clustering on unscaled eigenvectors with post-PCA spherical normalization \citep{NgJordanWeiss2001}, spectral clustering on scaled eigenvectors with post-PCA spherical normalization \citep{agterberg2025joint}, spectral clustering on scaled eigenvectors with post-PCA eigenvector normalization \citep{Jin2015}, and SCORE+ \citep{jin2022improvements}. All methods are conducted with the true number of clusters $K$. For each spectral clustering algorithm, the $K$-means algorithm is run with $10$ starting points and the result with smallest within-cluster variance is reported. 

As reported in Table \ref{tab:wdbc_clus}, all spectral clustering methods achieve competitive performance, reflecting the moderate dimensionality of these benchmark datasets relative to their sample sizes. Nevertheless, Algorithm \ref{alg:initial} consistently attains the lowest or nearly the lowest clustering error across all the datasets, demonstrating that Algorithm \ref{alg:initial} yields a systematic improvement over existing spectral clustering procedures rather than an isolated gain on a particular dataset.

\begin{table}[ht]
\centering
{
\begin{tabular}{rrrrr}
\toprule
Data & WDBC & Dermatology & Ecoli & Levine \\
\midrule
($n;p;K$) & $(569;30;2)$ & $(366;33;6)$ & $(336;7;8)$ & $(1147;32;3)$ \\
\midrule
Algorithm 1       & 0.180 & 0.262 & 0.283 & 0.267  \\
Unscaled eigenvector     & 0.200 & 0.270 & 0.467 & 0.341  \\
vanilla SC               & 0.188 & 0.268 & 0.283 & 0.468  \\
Spherical Normalization  & 0.183 & 0.267 & 0.458 & 0.455  \\
Eigenvector Normalization& 0.185 & 0.268 & 0.640 & 0.478  \\
SCORE+                     & 0.185 & 0.269 & 0.610 & 0.474  \\
\bottomrule
\end{tabular}
}
\caption{Clustering errors of Algorithm \ref{alg:initial} and competing methods on the WDBC, Dermatology, Ecoli, and Levine datasets. Algorithm \ref{alg:initial} outperforms the competing methods with smallest clustering error.}
\label{tab:wdbc_clus}
\end{table}

These empirical findings are well aligned with the theoretical analysis in Section \ref{sec:spectral}. The consistent improvement suggests that self-normalization effectively suppresses the influence of unstable spectral directions while preserving the dominant between-cluster structure, thereby producing more reliable spectral embeddings for subsequent clustering. 
The real-data experiments demonstrate that the theoretical advantages of self-normalization translate into meaningful and robust improvements in practical clustering performance.

\begin{table}[!h]
\centering
{
\begin{tabular}{rrrrr}
\toprule
Data & WDBC & Dermatology & Ecoli & Levine \\
\midrule
Algorithm 1  & 0.67 & 0.72 & 0.86 & 0.72 \\ 
vanilla SC & 0.65 & 0.67 & 0.69 & 0.55 \\ 
\bottomrule
\end{tabular}
}
\caption{Ratio of between-cluster sum of squares divided to total sum of squares for the $K$-means algorithm in Algorithm \ref{alg:initial} and vanilla spectral clustering on the WDBC, Dermatology, Ecoli, and Levine datasets. The self-normalization increases the ratio, so Algorithm \ref{alg:initial} outperforms vanilla spectral clustering.}
\label{tab:wdbc_ss}
\end{table}

Lastly, we report the between-cluster sum of squares divided by the total sum of squares for the $K$-means algorithm in Algorithm \ref{alg:initial} and vanilla spectral clustering in Table \ref{tab:wdbc_ss}. It is observed that the ratio increases after self-normalization, which reflects that self-normalization increases the signal-noise-ratio. This also shows the mechanism of self-normalization that it separates the eigen-directions induced by signals from those induced by errors.

\section{Conclusions and discussions}

The $g$-mixture model provides a useful semiparametric extension of the classical Gaussian mixture model by allowing unknown coordinate-wise monotone transformations in the latent space. Within this framework, our theory shows that spectral clustering remains statistically well justified beyond standard Gaussian-type settings. In particular, we establish the clustering consistency of spectral clustering with self-normalized embedding in Algorithm \ref{alg:initial} for fixed or slowly diverging dimension, thereby extending existing guarantees from Gaussian and sub-Gaussian mixture models to a broader class of non-Gaussian data-generating mechanisms. This result is especially relevant in applications where the observed variables exhibit skewness, heavy tails, or other marginal distortions that are not adequately captured by conventional Gaussian mixture assumptions.

Our analysis also sheds light on the role of spectral clustering in high-dimensional problems. Although spectral clustering is not always expected to be the final clustering procedure in such settings, both our theory indicates that it can still recover enough cluster structure to serve as a reliable first-stage estimator. This is important because many high-dimensional learning methods based on Gaussian mixtures or the $g$-mixture model are formulated through nonconvex optimization and are therefore sensitive to initialization. The simulation results show that spectral clustering can provide stable and informative starting values, even when its standalone clustering accuracy is not optimal. In this sense, spectral clustering plays a dual role: it is a theoretically analyzable clustering method in its own right, and it is also a practical and provably effective initialization device for more structured model-based procedures.

More broadly, our results suggest that the main value of spectral clustering in complex semiparametric settings lies not only in its direct clustering performance, but also in its ability to extract a low-dimensional representation that preserves the dominant between-cluster signal under weak distributional assumptions. This helps explain why spectral methods often remain competitive in practice even when the underlying model is substantially misspecified. At the same time, our theory is developed under the latent-transformation $g$-mixture model and therefore does not cover all possible forms of heterogeneity or dependence. Extending these guarantees to more general high-dimensional regimes, weaker structural assumptions, or settings with stronger within-cluster dependence would be an interesting direction for future work.

\section*{Appendix}

\renewcommand\thesubsection{S.\arabic{subsection}}
\renewcommand\theequation{S.\arabic{equation}}
\renewcommand\thelemma{S.\arabic{lemma}}
\setcounter{equation}{0}
\setcounter{lemma}{0}

\subsection{Proofs of main theorems} 
\label{sec:main_proof}

Throughout the appendix, let $L_4=\{2\log(4np)\}^{1/2}+\max_{1\le\ell\le p}(\max_k\mu_{k\ell}^*-\min_k\mu_{k\ell}^*)/\sigma_\ell^*$, $a_{n,p}=1-\Phi(L_4)$, and $\mathcal{D}_{n,p}=\{\bx\in\mathbb R^p:a_{n,p}\le\Phi(\{g_\ell^*(x_\ell)-\mu_{k\ell}^*\}/\sigma_\ell^*)\le1-a_{n,p}\text{ for each }k=1,\ldots,K\mbox{ and }\ell=1,\ldots,p\}$.
We first state the following lemmas whose proofs are given in Section \ref{sec:lem}.

\begin{lemma}\label{lem:g_approx}
Let $\bX_1,\ldots,\bX_n$ be independent observations from model~\eqref{eq:cesme}.
Suppose that $h_{\ell}=(g_{\ell}^*)^{-1}$ is locally absolutely continuous, $c_{\lambda}\le\lambda_k^*\le 1-c_{\lambda}$, $c_{\Sigma}^{-1}\le\sigma_{min}(\bSigma^*)\le\sigma_{max}(\bSigma^*)\le c_{\Sigma}$, $\max_{k,\ell}\int(\sigma_{\ell}^*)^{-1}\allowbreak\{\xi_{\ell}(y)\}^2\phi(\{y-\mu_{k\ell}^*\}/\sigma_{\ell}^*)dy<\infty$, we have
\begin{enumerate}[label=(\roman*)]
\item for any $\zeta>0$, $\mathbb{P}(\Vert n^{-1}\bX\bX^{\top}-n^{-1}\mathbb{E}[\bX\bX^{\top}]\Vert_2>\zeta)
\le p^2\max_{i,\ell}\mathbb{E}[X_{i\ell}^4]/(n\zeta^2)$;
\item for each $i$ and $k\ne k'$,
\begin{align*}
\Vert\mathbb{E}[\bX_i|z_i^*=k]-\mathbb{E}[\bX_i|z_i^*=k']\Vert_2\ge c_{\Sigma}^{-1/2}R_{\min}L_{\bg}.
\end{align*}
where $L_{\bg}=\inf_{\ell\in\mathcal{I}}\inf_{x\in J_{\ell}}H_{\ell}'(x)\}$, $H_{\ell}(x)=(\sigma_{\ell}^*)^{-1}\int_{-\infty}^{\infty}h_{\ell}(u)\phi((u-x)/\sigma_{\ell}^*)du$, $h_{\ell}$ is the inverse function of $g_{\ell}^*$, $J_{\ell}=[\min_k\mu_{k\ell}^*,\max_k\mu_{k\ell}^*]$, and $\mathcal{I}=\{\ell=1,\dots,p:\mu_{k\ell}^*\ne\mu_{k'\ell}^*\mbox{ for some }k\ne k'\}$;
\item for each $i$ and $k$, $0<\sigma_{min}(\cov(\bX_i|z_i^*=k))\le\sigma_{max}(\cov(\bX_i|z_i^*=k))\le c_{\Sigma}U_{\bg}^2$,
where $U_{\bg}^2=\max_{k,\ell}\int(\sigma_{\ell}^*)^{-1}\{h_{\ell}'(y)\}^2\phi(\{y-\mu_{k\ell}^*\}/\sigma_{\ell}^*)dy$;
\item for each $i$ and $k\ne k'$,
\begin{align*}
\Vert\{\cov(\bX_i|z_i^*=k)\}^{-1/2}(\mathbb{E}[\bX_i|z_i^*=k]-\mathbb{E}[\bX_i|z_i^*=k'])\Vert_2\ge C_{\bg}^{-1}c_{\Sigma}^{-1}R_{\min},
\end{align*}
where $C_{\bg}=U_{\bg}/L_{\bg}$.
\end{enumerate}

\end{lemma}

\begin{lemma}\label{lem:kernel_var}
Suppose that $\mathcal K$ satisfies Condition~\ref{cond:kernel}.
Let $\bX_1$ and $\bX_2$ be independent conditional on their labels $z_1^*$ and $z_2^*$, and assume that
$v_X=\max_{1\le k\le K}\operatorname{tr}\{\operatorname{Cov}(\bX_i| z_i^*=k)\}<\infty$.
Then, for every $k,\ell$,
\begin{align*}
\operatorname{Var}\{\mathcal K(\bX_1,\bX_2)
 | z_1^*=k,z_2^*=\ell\}
&\le 2\rho^2c_\Sigma U_{\bg}^2,\\
\operatorname{Var}\{\mathcal K(\bX_1,\bX_1)
 | z_1^*=k\}
&\le 4\rho^2c_\Sigma U_{\bg}^2.
\end{align*}
\end{lemma}

\begin{lemma}\label{lem:kernel}
Let $\bA^{\mathcal K}=\{\mathcal K(\bX_i,\bX_j)\}_{i,j=1}^n$ and $\bZ_{ik}=\{1(z_i^*=k)\}_{i,k=1}^{n,K}$. Define
$\bP_{\mathcal K}=\bZ\bB\bZ^\top$,
$\delta_k=\mathbb E\{\mathcal K(\bX_i,\bX_i)| z_i^*=k\}-B_{kk}$, 
$d_{\mathcal K}=\max_k|\delta_k|$.
For any $\zeta\in(0,1)$, conditional on $\bz^*$, with probability at least $1-\zeta$,
$$
\|\bA^{\mathcal K}-\bP_{\mathcal K}\|_2
\le
\frac{2\rho c_\Sigma^{1/2} U_{\bg}}{\zeta^{1/2}}+d_{\mathcal K}.
$$
\end{lemma}

\begin{lemma}\label{lem:tail}
Define $\mathcal{X}_n=\bigcap_{i=1}^n\{\bX_i\in\mathcal{D}_{n,p}\}$.
Under the choice of $a_{n,p}$ specified above, $\mathbb{P}(\mathcal{X}_n^c)\le n^{-1}/2$.
\end{lemma}

\begin{lemma}\label{lem:ghat}
Under Conditions of Theorem \ref{thm:kernel_spectral}, using the region $\mathcal{D}_{n,p}$ specified above, with probability at least $1-O(n^{-1})$,
$\Vert\widehat{\bg}^{(0)}-\bar{\bg}^{(0)}\Vert_{\infty,s}\le 3c_{\lambda}^{-2}\{\phi(\Phi^{-1}(a_{n,p}))\}^{-1}\mathcal{L}(\wh{\bz}^{(0)})+ \varepsilon_{\bg}(n,p,s,n^{-1})$,
where $\varepsilon_{\bg}(n,p,s,\zeta)=2\widetilde{\eta}[K^2\allowbreak\log\{nK^2(9s+1)/\zeta\}]^{1/2}n^{-1/2}$.
\end{lemma}

\begin{proof}[Proof of Theorem \ref{thm:kernel_spectral}]

In this paper, we work with the observed data matrix, rather than transferring the clustering guarantee directly from the latent Gaussian observations. Define $\bP_X=(\mathbb{E}[\bX_1| z_1^*],\ldots,\mathbb{E}[\bX_n| z_n^*])$ and $\bE_X=\bX-\bP_X$. Conditional on $\bz^*$, the columns of $\bE_X$ are independent and have mean zero. Write $
\Delta_X=\min_{k\ne k'}\|\mathbb{E}[\bX_i| z_i^*=k]-\mathbb{E}[\bX_i| z_i^*=k']\|_2$, $n_k=\sum_{i=1}^n1\{z_i^*=k\}$, and $n_{\min}=\min_{1\le k\le K}n_k$. By  Lemma~\ref{lem:g_approx} (ii), we have $\Delta_X^2\ge c_\Sigma^{-1}L_{\bg}^2R_{\min}^2$.

Let $\wh\bP=\wh\bU\wh\bD\wh\bV^\top$ be a best rank-$K$ approximation of $\bX$, where $\wh\bD=\diag(\wh d_1,\ldots,\wh d_K)$ and $\wh d_1\ge\cdots\ge\wh d_K\ge0$. For $\wh d_1>0$, define $\wh\bA_{\alpha}=(\bX\bX^\top+\alpha\wh d_1^2\bI_p)^{-1/2}$. Then, we have $\wh\bA_{\alpha}\wh\bP=\wh\bU(\wh\bD^2+\alpha\wh d_1^2\bI)^{-1/2}\wh\bD\wh\bV^\top$. Since $\wh\bU$ has orthonormal columns, applying $K$-means to the columns of $\wh\bA_{\alpha}\wh\bP$ is equivalent to applying
$K$-means to the embedding in Algorithm~\ref{alg:initial}.

We first establish a deterministic comparison with the cluster-constant matrix $\wh\bA_{\alpha}\bP_X$. Define $\mathcal{E}_\alpha=\|\wh\bA_{\alpha}(\wh\bP-\bP_X)\|_{\mathbb{F}}^2$ and $\Delta_{X,\alpha}=\min_{k\ne k'}\|\wh\bA_{\alpha}(\mathbb{E}[\bX_i| z_i^*=k]-\mathbb{E}[\bX_i| z_i^*=k'])\|_2$. Since $\operatorname{rank}(\bP_X)\le K$, the optimality of the truncated singular value decomposition gives that $\|\bX-\wh\bP\|_2\le \|\bX-\bP_X\|_2=\|\bE_X\|_2$. Consequently, we have $\|\wh\bP-\bP_X\|_2\le 2\|\bE_X\|_2$. Together with that $\operatorname{rank}(\wh\bP-\bP_X)\le2K$, this yields $\|\wh\bP-\bP_X\|_{\mathbb{F}}^2\le8K\|\bE_X\|_2^2$. Thus, we have $\mathcal{E}_\alpha\le8K\lambda_{\max}^2(\wh\bA_{\alpha})\|\bE_X\|_2^2$ and $\Delta_{X,\alpha}^2\ge\lambda_{\min}^2(\wh\bA_{\alpha})\Delta_X^2$. By the definition of $\wh\bA_{\alpha}$, for $\alpha > d_{K}^2/d_1^2$, we have $\lambda_{\max}^2(\wh\bA_{\alpha})/\lambda_{\min}^2(\wh\bA_{\alpha})\le2$, and hence
\begin{align}\label{eq:observed_embedding_comparison}
\frac{\mathcal{E}_\alpha}{\Delta_{X,\alpha}^2}\le16K\frac{\|\bE_X\|_2^2}{\Delta_X^2}.
\end{align}

Let $\wh\bC$ be the matrix whose $i$th column is the fitted center assigned to the $i$th column of $\wh\bA_{\alpha}\wh\bP$. Since $\wh\bA_{\alpha}\bP_X$ has at most $K$ distinct columns, the optimality of $K$-means implies $\|\wh\bC-\wh\bA_{\alpha}\wh\bP\|_{\mathbb{F}}^2\le\mathcal{E}_\alpha$. The triangle inequality then gives $\|\wh\bC-\wh\bA_{\alpha}\bP_X\|_{\mathbb{F}}^2\le4\mathcal{E}_\alpha$. Define $\mathcal{N}=\{i:\|\wh\bC_{\cdot i}-\bA_{\alpha}\mathbb{E}[\bX_i| z_i^*]\|_2\ge \Delta_{X,\alpha}/2\}$. Since $\|\widehat{\bC}-\widehat{\bA}_\alpha\bP_X\|_{\mathbb{F}}^2\le4\mathcal E_\alpha$ and $R_{\min}\ge 8c_{\lambda}^{-1/2}Kc_{\Sigma}C_{\bg}$, we have
$$
|\mathcal N|
\le\frac{16\mathcal E_\alpha}{\Delta_{X,\alpha}^2}\le nc_{\lambda}/2.
$$
For each $k$, the set $\mathcal N_k=\{i:z_i^*=k,\ i\notin\mathcal N\}$ is then nonempty. Moreover, observations in $\mathcal N_k$ and $\mathcal N_{k'}$ for $k\ne k'$ cannot share a fitted center: otherwise, the triangle inequality would give a distance strictly less than $\Delta_{X,\alpha}$ between their corresponding transformed population centers. Thus, the sets of fitted labels assigned to $\mathcal N_1,\ldots,\mathcal N_K$ are nonempty and pairwise disjoint. Since there are at most $K$ fitted labels, each set contains exactly one label. Consequently, there exists a permutation $\Pi$ such that
$\widehat z_i=\Pi(z_i^*)$ for every $i\notin\mathcal N$, and
$$
\mathcal L(\widehat{\bz})
\le\frac{|\mathcal N|}{n}
\le\frac{16\mathcal E_\alpha}
{n\Delta_{X,\alpha}^2}\le\frac{16Kc_{\Sigma}^2C_{\bg}^2(n+p)}{nR_{\min}^2}.
$$

Then, denoting 
$$\mathcal{A}=\bigg\{\min_{\Pi\in\mathcal{S}_K}\frac1n\sum_{i=1}^n1(\widehat{z}_i\ne \Pi z_i^*) \le \frac{16Kc_{\Sigma}^2C_{\bg}^2(n+p)}{nR_{\min}^2}\bigg\},$$ 
we have
\begin{align*}
\mathbb{E}[\mathcal{L}(\wh\bz)]
=&~\mathbb{E}\bigg[\min_{\Pi\in\mathcal{S}_K}\frac1n\sum_{i=1}^n1(\widehat{z}_i\ne \Pi z_i^*)\bigg]\\
\le& ~\mathbb{E}\bigg[\min_{\Pi\in\mathcal{S}_K}\frac1n\sum_{i=1}^n1(\widehat{z}_i\ne \Pi z_i^*)| \mathcal{A}\bigg]\mathbb{P}\{\mathcal{A}\}
+\mathbb{E}\bigg[\min_{\Pi\in\mathcal{S}_K}\frac1n\sum_{i=1}^n1(\widehat{z}_i\ne \Pi z_i^*)| \mathcal{A}^c\bigg]\mathbb{P}\{\mathcal{A}^c\}\\
\le&~ \frac{16Kc_{\Sigma}^2C_{\bg}^2(n+p)}{nR_{\min}^2}+\mathbb{P}\{\mathcal{A}^c\}\\
\le&~ \frac{16Kc_{\Sigma}^2C_{\bg}^2(n+p)}{nR_{\min}^2}+\frac Cn,
\end{align*}
for some positive constant $C$.
The conclusion follows.

\end{proof}

\begin{proof}[Proof of Corollary \ref{cor:kernel_spectral}]

First, for any $\zeta\in(0,1)$, we consider a positive constant $\epsilon_\zeta$ depending on $\zeta$ and the event
$
\|\bA^{\mathcal K}-\bP_{\mathcal K}\|_2
\le\epsilon_\zeta.
$
Denote
$\bP_{\mathcal K}=\bV\bS\bV^\top$ and
$\bS=\operatorname{diag}(s_1,\ldots,s_K)$,
and let $\wh\bV$ contain the leading $K$ eigenvectors
of $\bA^{\mathcal K}$, with corresponding eigenvalues
$\wh s_1\ge\cdots\ge\wh s_K$.
Weyl's inequality gives $\wh s_K\ge s_K-\epsilon_\zeta\ge s_K/2$ and 
$\wh s_1\le s_1+\epsilon_\zeta\le3s_1/2$.
Let $\bE_{\mathcal K}=\bA^{\mathcal K}-\bP_{\mathcal K}$.
Projecting the empirical eigenvector equation onto
the orthogonal complement of $\operatorname{col}(\bV)$
gives
$(\bI-\bV\bV^\top)\wh\bV\operatorname{diag}(\wh s_1,\ldots,\wh s_K)=(\bI-\bV\bV^\top)\bE_{\mathcal K}\wh\bV$.
Then, we have
$\|(\bI-\bV\bV^\top)\wh\bV\|_{\mathbb{F}}\le 2\sqrt K\,\epsilon_\zeta/s_K$.
By orthogonal Procrustes alignment, there exists
$\bm{O}\in\mathbb R^{K\times K}$ with
$\bm{O}^\top\bm{O}=\bI_K$ such that
$
\|\wh\bV-\bV\bm{O}\|_{\mathbb{F}}^2
\le2\|(\bI-\bV\bV^\top)\wh\bV\|_{\mathbb{F}}^2
\le8K\epsilon_\zeta^2/s_K^2.
$

Recall that the empirical self-normalization weights is
$
\wh\bW^{\text{SN}}
=\operatorname{diag}\{
\sqrt{\wh s_j/(\wh s_j+\alpha\wh s_1)}\}_{j=1}^K$.
Then, we have $\|\wh\bW^{\text{SN}}\|_2\le1$ and $\sigma_{\min}^2(\wh\bW^{\text{SN}})\ge
(1+3\alpha\kappa_{\mathcal K})^{-1}$.
Let
$\bm{N}=\operatorname{diag}(n_1,\ldots,n_K)$.
Since $\bB$ is positive definite,
$\operatorname{col}(\bV)=\operatorname{col}(\bZ)$,
and hence $\bV=\bZ\bm{N}^{-1/2}\bR$
for some orthogonal matrix $\bR$.
For observations $i,j$ from different clusters $k,\ell$, we have
$\|\bV_{i\cdot}-\bV_{j\cdot}\|_2^2=
n_k^{-1}+n_\ell^{-1}$.
Then, the minimum squared separation $\Delta_0^2$ between distinct rows of $\bF_0$ satisfies $\Delta_0^2\ge 2/\{n_{\max}(1+3\alpha\kappa_{\mathcal K})\}$.

Let $\wh\bC$ denote the fitted centroid matrix from
a global $K$-means solution on $\wh\bF$.
Since $\bF_0$ has $K$ distinct rows, $\|\wh\bF-\wh\bC\|_{\mathbb{F}}^2
\le
\|\wh\bF-\bF_0\|_{\mathbb{F}}^2$. 
Thus, $\|\wh\bC-\bF_0\|_{\mathbb{F}}^2
\le4\|\wh\bF-\bF_0\|_{\mathbb{F}}^2$.
Define $\mathcal B=
\left\{
i:\|\wh\bC_{i\cdot}-(\bF_0)_{i\cdot}\|_2
\ge\Delta_0/2
\right\}$.
Then
\begin{align*}
|\mathcal B|
\le\frac{4\|\wh\bC-\bF_0\|_{\mathbb{F}}^2}{\Delta_0^2}
\le\frac{16\|\wh\bF-\bF_0\|_{\mathbb{F}}^2}{\Delta_0^2}
\le64K n_{\max}(1+3\alpha\kappa_{\mathcal K})
\frac{\epsilon_\zeta^2}{s_K^2}
<n_{\min}.
\end{align*}
Each true cluster therefore contains an observation
outside $\mathcal B$. Observations outside $\mathcal B$
from different true clusters cannot share a fitted
centroid, by the definition of $\Delta_0$.
Since there are $K$ true clusters and at most $K$
fitted centroids, these centroids can be matched
bijectively to the true clusters. Under this matching,
all observations outside $\mathcal B$ are correctly
classified. Hence, $\mathcal L(\wh\bz)\le|\mathcal B|/n$, 
which proves the result.

\end{proof}

\subsection{Proofs of initialization guarantee} 
\label{sec:proof_initial}

\begin{proof}[Proof of Proposition \ref{cor:initial_0}]

By Lemma \ref{lem:ghat} and Theorem \ref{thm:kernel_spectral}, with probability at least $1-O(n^{-1})$, 
\begin{align}
\Vert\widehat{\bg}^{(0)}-\bg^*\Vert_{\infty,s}^2\le 6c_{\lambda}^{-2}\mathcal{L}(\widehat{\bz}^{(0)})+2\varepsilon_{\bg}^2(n,p,s,n^{-1})\le R_{\min}^2/32\label{eqn:sc1}
\end{align} 
for sufficiently large $n$ such that $\varepsilon_{\bg}^2(n,p,s,n^{-1})\le R_{\min}^2/128$.
Note that in Theorem \ref{thm:kernel_spectral}, the minimization over permutations $\Pi $ is introduced to avoid errors from label switching. Without loss of generality, we assume that 
$n^{-1}\sum_{i=1}^n1(\widehat{z}_i\ne \Pi z_i^*)$ and $\Vert\wh\bmu_{k}-\bmu_{\Pi k^*}\Vert_{\infty}^2$ is minimized at 
$\Pi \in\mathcal{S}_K$ satisfying that $\Pi k=k$ for each $k=1,\dots,K$, and thus the permutation $\Pi$ is omitted in the rest of proof. Then, by Theorem \ref{thm:kernel_spectral}, with probability at least $1-O(n^{-1})$,
\begin{align*}
&~\Vert\wh\bmu_{k}^{(0)}-\bmu_{k}^*\Vert_{2,s}^2\\
\le&~ \left\Vert\frac{\sum_i 1(\wh z_{i}^{(0)}=k)\wh{\bg}^{(0)}(X_i)}{\sum_i 1(\wh z_{i}^{(0)}=k)}-\bmu_{k}^*\right\Vert_{2}^2\\
\le&~ \left\Vert\frac{\sum_i 1(\wh z_{i}^{(0)}=k)\wh{\bg}^{(0)}(X_i)}{\sum_i 1(\wh z_{i}^{(0)}=k)}-\frac{\sum_i 1(\wh z_{i}^{(0)}=k)\bg^*(X_i)}{\sum_i 1(\wh z_{i}^{(0)}=k)}\right\Vert_{2}^2
+\left\Vert\frac{\sum_i 1(\wh z_{i}^{(0)}=k)\bg^*(X_i)}{\sum_i 1(\wh z_{i}^{(0)}=k)}-\bmu_{k}^*\right\Vert_{2}^2\\
\le&~ \frac{16K^2C_{\bg}^2c_{\Sigma}^2}{R_{\min}^2}.
\end{align*} 
Hence, for $R_{\min}\ge8c_\lambda^{-1/2}KC_{\bg}c_{\Sigma}$, we have, with probability at least $1-O(n^{-1})$,
\begin{align}
\Vert\wh\bmu_{k}^{(0)}-\bmu_{k}^*\Vert_{2,s}^2\le R_{\min}^2/16.\label{eqn:sc2}
\end{align}
Then, the pooled sample covariance $
\wh\bSigma^{(0)}=n^{-1}\sum_{k=1}^K\sum_{\wh z_i=k}\{\wh \bg^{(0)}(\bX_i)-\wh\bmu_k^{(0)}\}^{\top}\{\wh \bg^{(0)}(\bX_i)-\wh\bmu_k^{(0)}\}^{\top}$ satisfies that, with probability at least $1-O(n^{-1})$, \allowdisplaybreaks
\begin{align*}
&\;\Vert \wh\bSigma^{(0)}-\bSigma^*\Vert_{2,s}^2\\
= &\;\;\bigg\Vert \frac{1}{n}\sum_{k=1}^K\sum_{\wh z_i=k}\{\wh \bg^{(0)}(\bX_i)-\wh\bmu_k^{(0)}\}\{\wh \bg^{(0)}(\bX_i)-\wh\bmu_k^{(0)}\}^{\top}-\bSigma^*\bigg\Vert_{2,s}^2\\
\le &\;\bigg\Vert \frac{1}{n}\sum_{k=1}^K\sum_{\wh z_i=k}\{\wh \bg^{(0)}(\bX_i)-\wh\bmu_k^{(0)}\}\{\wh \bg^{(0)}(\bX_i)-\wh\bmu_k^{(0)}\}^{\top}\\
&\;-\frac{1}{n}\sum_{k=1}^K\sum_{\wh z_i=k}\{\bg^*(\bX_i)-\wh\bmu_k^{(0)}\}\{\bg^*(\bX_i)-\wh\bmu_k^{(0)}\}^{\top}\bigg\Vert_{2,s}^2\\
&\;+\bigg\Vert \frac{1}{n}\sum_{k=1}^K\sum_{\wh z_i=k}\{\bg^*(\bX_i)-\wh\bmu_k^{(0)}\}\{\bg^*(\bX_i)-\wh\bmu_k^{(0)}\}^{\top}-\bSigma^*\bigg\Vert_{2,s}^2\\
\stackrel{(i)}{\leq} &\;\bigg\Vert \frac{2}{n}\sum_{i=1}^n(\bg^*(\bX_i)-\bmu_{z_i^*}^*)(\wh\bmu_{\wh z_i}^{(0)}-\bmu_{\wh z_i}^*)^{\top}\bigg\Vert_{2,s}^2\\
&\;+\bigg\Vert \frac{1}{n}\sum_{i=1}^n(\bg^*(\bX_i)-\bmu_{z_i^*}^*)(\bg^*(\bX_i)-\bmu_{z_i^*}^*)^{\top}-\bSigma^*\bigg\Vert_{2,s}^2+\frac{R_{\min}^2}{32}\\
\stackrel{(ii)}{\leq} &\;2\bigg\Vert \frac{1}{n}\sum_{i=1}^n(\wh\bmu_{\wh z_i}^{(0)}-\bmu_{\wh z_i}^*)(\wh\bmu_{\wh z_i}^{(0)}-\bmu_{\wh z_i}^*)^{\top}\bigg\Vert_{2,s}
\bigg\Vert \frac{1}{n}\sum_{i=1}^n(\bg^*(\bX_i)-\bmu_{z_i^*}^*)(\bg^*(\bX_i)-\bmu_{z_i^*}^*)^{\top}\bigg\Vert_{2,s}\\
&\;+\bigg\Vert \frac{1}{n}\sum_{i=1}^n(\bg^*(\bX_i)-\bmu_{z_i^*}^*)(\bg^*(\bX_i)-\bmu_{z_i^*}^*)^{\top}-\bSigma^*\bigg\Vert_{2,s}^2
+\frac{R_{\min}^2}{32}\\
\stackrel{(iii)}{\leq} &\; 2c_{\Sigma}\bigg\{1+\sqrt{\frac{s\log(p)}{n}}\bigg\}\Vert\wh\bmu_{k}^{(0)}-\bmu_{k}^*\Vert_{2,s}^2
+\frac{c_{\Sigma}^2s\log(p)}{n}+\frac{R_{\min}^2}{32}\\
\le&\; \frac{R_{\min}^2}{16},
\end{align*}
where $(i)$, $(ii)$, and $(iii)$ follow from \eqref{eqn:sc1}, Cauchy-Schwartz inequality, and the concentration inequality of $\bg^*(\bX_i)$, respectively, and the last inequality follows from \eqref{eqn:sc2}, that $c_{\Sigma}^2s\log(p)/n<R_{\min}^2/64$ for sufficiently large $n$, and that $R_{\min}^2\ge64c_{\Sigma}$.
The conclusion follows by letting $C_0=4 c_{\lambda}^{-1/2}c_{\Sigma}$.
\end{proof}

\begin{proof}[Proof of Proposition \ref{cor:initial_1}]

	Note that in Theorem \ref{thm:kernel_spectral}, we introduce minimization over permutations $\Pi $ to avoid errors from label switching. Without loss of generality, we assume that 
	$n^{-1}\sum_{i=1}^n1(\widehat{z}_i\ne \Pi z_i^*)$ and $\Vert\wh\bmu_{k}-\bmu_{\Pi k^*}\Vert_{\infty}^2$ is minimized at 
	$\Pi \in\mathcal{S}_K$ satisfying that $\Pi k=k$ for each $k=1,\dots,K$, and thus the permutation $\Pi$ is omitted in the rest of proof. Then, by \eqref{eqn:sc2}, with probability at least $1-O(n^{-1})$,
	\begin{align*}
		\Vert\wh\bmu_{k}^{(0)}-\bmu_{k}^*\Vert_{2}^2
		\le \frac{16C_{\bg}^2c_{\Sigma}^2(n+p)}{nR_{\min}^2}.
	\end{align*} 
    Since $R_{\min}\ge8c_\lambda^{-1/2}KC_{\bg}c_{\Sigma}$, we have $\Vert\wh\bmu_{k}^{(0)}-\bmu_{k}^*\Vert_{2}^2\le R_{\min}^2/64$.
	Also, since $\min_{j\ne k}\Vert\bmu_k^*-\bmu_j^*\Vert_2\ge8c_\lambda^{-1/2}KC_{\bg}c_{\Sigma}$, we have 
$\Vert\wh\bmu_{k}^{(0)}-\bmu_{k}^*\Vert_{2}^2\le \min_{j\ne k}\Vert\bmu_k^*-\bmu_j^*\Vert_2^2/64$. 
	The conclusion follows.
\end{proof}

\begin{proof}[Proof of Proposition \ref{cor:initial_2}]
	
	Note that in Theorem \ref{thm:kernel_spectral}, we introduce minimization over permutations $\Pi $ to avoid errors from label switching. Without loss of generality, we assume that 
	$n^{-1}\sum_{i=1}^n1(\widehat{z}_i\ne \Pi z_i^*)$ and $\Vert\wh\bmu_{k}-\bmu_{\Pi k^*}\Vert_{\infty}^2$ is minimized at 
	$\Pi \in\mathcal{S}_2$ satisfying that $\Pi k=k$ for each $k=1,2$, and thus the permutation $\Pi$ is omitted in the rest of proof. Then, by \eqref{eqn:sc2},  with probability at least $1-O(n^{-1})$,
	\begin{align*}
		\Vert\wh\bmu_{k}^{(0)}-\bmu_{k}^*\Vert_{2}^2
		\le \frac{16C_{\bg}^2c_{\Sigma}^2(n+p)}{nR_{\min}^2}.
	\end{align*} 
	The conclusion follows from Davis-Kahan Theorem \citep{yu2014useful}.
\end{proof}

\subsection{Proofs of technical lemmas}
\label{sec:lem}

\begin{proof}[Proof of Lemma \ref{lem:g_approx}]

\begin{enumerate}[label=(\roman*)]
\item 
Recall that for some $\delta>0$, $$\max_{1\leq \ell\leq p}\sum_{k=1}^K\lambda_k^*\int_{-\infty}^{\infty} |x|^{4+\delta}
\frac{1}{\sigma_\ell^*}\allowbreak\phi\left(\frac{g_\ell^*(x)-\mu_{k\ell}^*}{\sigma_\ell^*}\right)d g_\ell^*(x) \leq M_{4+\delta}<\infty,$$ where $\phi$ is {\it pdf} of univariate standard Gaussian distribution and $dg_{\ell}^*(x)$ the Lebesgue–Stieltjes measure induced by $g_{\ell}^*(x)$. Also, recall that for each $i=1,\dots,n$, conditional on $z_i^*$, $\bg^*(\bX_i)\sim\mathcal{N}(\bmu_{z_i^*}^*,\bSigma^*)$ and $g_{\ell}^*(X_{i\ell})\sim\mathcal{N}(\mu_{z_i^*\ell}^*,(\sigma_{\ell}^*)^2)$ for each $\ell=1,\dots,p$.  Then, we have $\max_{i,\ell}\mathbb{E}[|X_{i\ell}|^{4+\delta}]<\infty$. 
Also, we have
\begin{align*}
\mathbb{E}[\Vert n^{-1}\bX\bX^{\top}-n^{-1}\mathbb{E}[\bX\bX^{\top}]\Vert_{\mathbb{F}}^2]
\le n^{-1}\sum_{\ell=1}^p\sum_{\ell'=1}^p\mathbb{E}[X_{i\ell}^2X_{i\ell'}^2]
\le n^{-1}p^2\max_{i,\ell}\mathbb{E}[X_{i\ell}^4].
\end{align*}
Then, for any $\zeta>0$, we have
\begin{align*}
&~\mathbb{P}(\Vert n^{-1}\bX\bX^{\top}-n^{-1}\mathbb{E}[\bX\bX^{\top}]\Vert_2>\zeta)
\le\frac{p^2\max_{i,\ell}\mathbb{E}[X_{i\ell}^4]}{n\zeta^2}.
\end{align*}

\item
Then, for each $\ell=1,\dots,p$, we define $J_{\ell}=[\min_k\mu_{k\ell}^*,\max_k\mu_{k\ell}^*]$. 
Recall that $h_{\ell}$ is the inverse function of $g_{\ell}^*$ for each $\ell=1,\dots,p$. For each $\ell=1,\dots,p$, we define
\begin{align*}
H_{\ell}(x)=\frac{1}{\sigma_{\ell}^*}\int_{-\infty}^{\infty}h_{\ell}(u)\phi((u-x)/\sigma_{\ell}^*)du,
\end{align*}
where $\phi(\cdot)$ is the {\it pdf} of univariate standard Gaussian distribution. 
We first justify differentiation and continuity of the derivative. Fix a component mean $a=\mu_{k\ell}^*$, and let $p_a$ denote the density of $\mathcal{N}(a,(\sigma_\ell^*)^2)$. The moment assumption and $\lambda_k^*>0$ imply that $h_\ell\in L^2(p_a)$. Moreover,
\begin{align*}
\frac{p_x(u)}{p_a(u)}=\exp\left\{\frac{(x-a)(u-a)}{(\sigma_\ell^*)^2}-\frac{(x-a)^2}{2(\sigma_\ell^*)^2}\right\}
\end{align*}
is continuously differentiable in $x$ as an $L^2(p_a)$-valued function. Differentiating the corresponding $L^2(p_a)$ pairing therefore shows that $H_\ell$ is continuously differentiable, with
\begin{align*}
H_{\ell}'(x)=\int_{-\infty}^{\infty}\frac{u-x}{(\sigma_{\ell}^*)^3}h_{\ell}(u)\phi((u-x)/\sigma_{\ell}^*)du
=\frac{1}{\sigma_{\ell}^*}\mathbb{E}[Wh_{\ell}(x+\sigma_{\ell}^*W)],
\end{align*}
where $W\sim\mathcal{N}(0,1)$. Let $W'$ be an independent copy of $W$. Then, we have
\begin{align*}
H_\ell'(x)=\frac{1}{2\sigma_\ell^*}\mathbb{E}\left[(W-W')\{h_\ell(x+\sigma_\ell^*W)-h_\ell(x+\sigma_\ell^*W')\}\right]>0,
\end{align*}
where strict positivity follows from strict monotonicity of $h_\ell$ and that $\mathbb{P}(W=W')=0$. Since $J_{\ell}$ is compact and not empty, we have $\inf_{x\in J_{\ell}}H_{\ell}'(x)>0$. Then, for each $i$ and $k\ne k'$, if $\mu_{k\ell}^*\ne\mu_{k'\ell}^*$,
\begin{align*}
|\mathbb{E}[X_{i\ell}|z_i^*=k]-\mathbb{E}[X_{i\ell}|z_i^*=k']|
=|H_{\ell}(\mu_{k\ell}^*)-H_{\ell}(\mu_{k'\ell}^*)|
\ge \inf_{x\in J_{\ell}}H_{\ell}'(x)|\mu_{k\ell}^*-\mu_{k'\ell}^*|,
\end{align*}
and 
\begin{align*}
\Vert\mathbb{E}[\bX_i|z_i^*=k]-\mathbb{E}[\bX_i|z_i^*=k']\Vert_2\ge c_{\Sigma}^{-1/2}R_{\min}\inf_{\ell\in\mathcal{I}}\inf_{x\in J_{\ell}}H_{\ell}'(x),
\end{align*}
where $\mathcal{I}=\{\ell=1,\dots,p:\mu_{k\ell}^*\ne\mu_{k'\ell}^*\mbox{ for some }k\ne k'\}$.
Since $R_{\min}=\min_{j\ne k}\Vert(\bSigma^*)^{-1/2}(\bmu_k^*-\bmu_j^*)\Vert_2>0$, $\bmu_k^*$ and $\bmu_{k'}^*$ are distinct for each $k\ne k'$, so $\mathcal{I}$ is not empty.
The conclusion follows by letting $L_{\bg}=\inf_{\ell\in\mathcal{I}}\inf_{x\in J_{\ell}}H_{\ell}'(x)\}$.

\item 
Since $\max_{i,\ell}\mathbb{E}[|X_{i\ell}|^{4+\delta}]<\infty$ with $\delta>0$, we have $
\mathbb{E}[h_{\ell}^2(\mu_{k\ell}^*+\sigma_{\ell}^*W)]<\infty$ for each $k$ and $\ell$ and  $W\sim\mathcal{N}(0,1)$.
Then, for each $i$, $k$, and $\ell$, we have
\begin{align*}
\mathbb{E}[\{g_{\ell}^*(X_{i\ell})-\mu_{k\ell}^*\}X_{i\ell}| z_i^*=k]=\sigma_{\ell}^*\mathbb{E}[Wh_{\ell}(x+\sigma_{\ell}^*W)]
\end{align*}
with $W\sim\mathcal{N}(0,1)$.
Since the covariance matrix of $(\bX^{\top},\bg^*(\bX)^{\top})^{\top}$ is non-negative definite, we have, for each $k$,
\begin{align*}
\cov(\bX_i|z_i^*=k)\succeq\Upsilon_k\bSigma^*\Upsilon_k\succ0,
\end{align*}
since $\bSigma^*\succ0$ and each diagonal entry of $\Upsilon_k=\diag(H_1'(\mu_{k1}),\dots,H_p'(\mu_{kp}))$ is positive.

Also, for each $k$ and any unit vector $\ba\in\mathbb{R}^p$, by Gaussian Poincaré inequality, 
\begin{align*}
\var(\ba^{\top}\bX_i|z_i^*=k)
\le \mathbb{E}[\ba^{\top}\nabla^{\top}\bh(\bY)\bSigma^*\nabla \bh(\bY)\ba]
\le c_{\Sigma}\sum_{\ell=1}^pa_{\ell}^2\mathbb{E}[\{\xi_{\ell}(Y_{\ell})\}^2]
\le c_{\Sigma}U_{\bg}^2,
\end{align*}
where $U_{\bg}^2=\max_{k,\ell}\int(\sigma_{\ell}^*)^{-1}\{\xi_{\ell}(y)\}^2\phi(\{y-\mu_{k\ell}^*\}/\sigma_{\ell}^*)dy$
and $\bY\sim\mathcal{N}(\bmu_k^*,\bSigma^*)$.
The conclusion follows.
\allowdisplaybreaks
\item By (ii) and (iii),
\begin{align*}
&\Vert\{\cov(\bX_i|z_i^*=k)\}^{-1/2}(\mathbb{E}[\bX_i|z_i^*=k]-\mathbb{E}[\{\bX_i|z_i^*=k'])\Vert_2\\
\ge&~\frac{\|\mathbb{E}[\bX_i|z_i^*=k]-\mathbb{E}[\{\bX_i|z_i^*=k']\|_2}{\{\sigma_{\max}(\cov(\bX_i|z_i^*=k))\}^{1/2}}\\
\ge&~\frac{c_\Sigma^{-1/2}L_{\bg}R_{\min}}{c_\Sigma^{1/2}U_{\bg}}\\
=&~C_{\bg}^{-1}c_\Sigma^{-1}R_{\min},
\end{align*}
where $C_{\bg}=U_{\bg}/L_{\bg}$.
The conclusion follows.

\end{enumerate}

\end{proof}

\allowdisplaybreaks

\begin{proof}[Proof of Lemma \ref{lem:kernel_var}]

In this proof, all expectations and variances below are conditional on the indicated labels $\bz^*$. Let $\bX_1'$ and $\bX_2'$
be independent copies of $\bX_1$ and $\bX_2$,
respectively. By the Efron--Stein inequality and the
Lipschitz property,
\begin{align*}
\operatorname{Var}\{\mathcal K(\bX_1,\bX_2)\}
&\le
\frac12\mathbb E\{
\mathcal K(\bX_1,\bX_2)
-\mathcal K(\bX_1',\bX_2)\}^2+
\frac12\mathbb E\{
\mathcal K(\bX_1,\bX_2)
-\mathcal K(\bX_1,\bX_2')\}^2\\
&\le
\frac{\rho^2}{2}
\left\{
\mathbb E\|\bX_1-\bX_1'\|_2^2+
\mathbb E\|\bX_2-\bX_2'\|_2^2
\right\}\\
&\le 2\rho^2v_X.
\end{align*}
Similarly, since $x\mapsto\mathcal K(x,x)$ is
$2\rho$-Lipschitz,
\begin{align*}
\operatorname{Var}\{\mathcal K(\bX_1,\bX_1)\}
&=
\frac12\mathbb E\{
\mathcal K(\bX_1,\bX_1)
-\mathcal K(\bX_1',\bX_1')\}^2
\le
2\rho^2\mathbb E\|\bX_1-\bX_1'\|_2^2
\le 4\rho^2v_X.
\end{align*}

For each coordinate, the Gaussian Poincare inequality
applied to $X_{i\ell}=h_\ell(Y_{i\ell})$ gives
$$
\operatorname{Var}(X_{i\ell}| z_i^*=k)
\le
(\sigma_\ell^*)^2
\mathbb E\{h_\ell'(Y_{i\ell})^2| z_i^*=k\}
\le c_\Sigma U_{\bg}^2.
$$
Summing over $\ell$ yields $v_X\le p c_\Sigma U_{\bg}^2$.

\end{proof}

\begin{proof}[Proof of Lemma \ref{lem:kernel}]

Let
$
\bM_{\mathcal K}
=
\mathbb E[\bA^{\mathcal K}|\bz^*].
$
Conditional independence of the observations gives
$
\bM_{\mathcal K}
=
\bP_{\mathcal K}
+
\operatorname{diag}(\delta_{z_1^*},\ldots,\delta_{z_n^*}).
$
Moreover, denoting $V_{\mathcal K}=\sum_{i,j=1}^n\operatorname{Var}\{\mathcal K(\bX_i,\bX_j)|\bz^*\}$, we have
$
\mathbb E[
\|\bA^{\mathcal K}-\bM_{\mathcal K}\|_{\mathbb{F}}^2
|\bz^*]
=
V_{\mathcal K}.
$
This identity does not require independence of the kernel entries. By Markov's inequality, we have
$\mathbb P\{\|\bA^{\mathcal K}-\bM_{\mathcal K}\|_2>\sqrt{V_{\mathcal K}/\zeta}|\bz^*\}\le\zeta$.
By triangle inequality, conditional on $\bz^*$, with probability at least $1-\zeta$,
$$
\|\bA^{\mathcal K}-\bP_{\mathcal K}\|_2
\le
\sqrt{\frac{V_{\mathcal K}}{\zeta}}+d_{\mathcal K}.
$$
Then, by Lemma~\ref{lem:kernel_var}, we have
$V_{\mathcal K}\le2\rho^2v_X n(n-1)+4\rho^2v_X n=2\rho^2v_X n(n+1)$,
where $v_X$ is defined in Lemma~\ref{lem:kernel_var}.
Finally, for independent observations $\bX,\bX'$
from component $k$, $
|\delta_k|=
\left|
\mathbb E\{\mathcal K(\bX,\bX)-\mathcal K(\bX,\bX')\}
\right|
\le\rho\,\mathbb E\|\bX-\bX'\|_2
\le \rho\sqrt{2v_X}$.
The conclusion follows.

\end{proof}

\begin{proof}[Proof of Lemma \ref{lem:tail}]
Recall that $g_{\ell}^*(X_{\ell})\sim \sum_{k=1}^K\lambda_k^*\mathcal{N}(\mu_{k\ell}^*,(\sigma_{\ell}^*)^2)$ for each $\ell=1,\dots,p$, where $g_{\ell}^*$ is the $\ell$th entry of $\bg^*$, $\mu_{k\ell}^*$ is the $\ell$th element of $\bmu_k^*$, and $(\sigma_{\ell}^*)^2$ is the $\ell$th diagonal entry of $\bSigma^*$. 
For each $k=1,\ldots,K$, conditional on $z_i^*=j$,
$$
\left|
\frac{g_\ell^*(X_{i\ell})-\mu_{k\ell}^*}{\sigma_\ell^*}
\right|
\le
\frac{|g_\ell^*(X_{i\ell})-\mu_{j\ell}^*|}{\sigma_\ell^*}
+
\frac{|\mu_{j\ell}^*-\mu_{k\ell}^*|}{\sigma_\ell^*}
\le \frac{|g_\ell^*(X_{i\ell})-\mu_{j\ell}^*|}{\sigma_\ell^*}+\max_{1\le\ell\le p}
\frac{\max_k\mu_{k\ell}^*-\min_k\mu_{k\ell}^*}
{\sigma_\ell^*}.
$$
Consequently, $|g_\ell^*(X_{i\ell})-\mu_{j\ell}^*|/\sigma_\ell^*\le \{2\log(4np/\zeta)\}^{1/2}$ implies that for each $k=1,\dots,K$,
$$
a_{n,p}
\le
\Phi\left(
\frac{g_\ell^*(X_{i\ell})-\mu_{k\ell}^*}{\sigma_\ell^*}
\right)
\le1-a_{n,p}.
$$
The Gaussian tail bound therefore gives
$$
\mathbb P\left\{
\max_{1\le k\le K}
\left|
\frac{g_\ell^*(X_{i\ell})-\mu_{k\ell}^*}{\sigma_\ell^*}
\right|>L_4
\,\middle|\,z_i^*=j
\right\}
\le 2e^{-\log(4np/\zeta)}
=\frac{\zeta}{2np}.
$$
Averaging over $j$ and applying the union bound over $i$
and $\ell$, we obtain
\begin{align*}
\mathbb P(\mathcal X_n^c)
&\le
\sum_{i=1}^n\sum_{\ell=1}^p
\mathbb P\left\{
\max_{1\le k\le K}
\left|
\frac{g_\ell^*(X_{i\ell})-\mu_{k\ell}^*}{\sigma_\ell^*}
\right|>L_4
\right\}
\le np\,\frac{\zeta}{2np}
=\frac{\zeta}{2}.
\end{align*}
The conclusion follows by letting $\zeta=n^{-1}$.
\end{proof}

\begin{proof}[Proof of Lemma \ref{lem:ghat}]

Recall that $\widehat{g}_{\ell}(x)=\wh\sigma_{\ell}\sum_{k=1}^K\wh\lambda_k\Phi^{-1}(\widehat{\psi}_{\ell k}(x))$ and $g_{\ell}^*(x)=\sigma_{\ell}^*\sum_{k=1}^K\lambda_k^*\allowbreak\Phi^{-1}(\psi_{\ell k}(x))$. 
Since $\sum_{k=1}^K\wh\lambda_k\Phi^{-1}(\widehat{\psi}_{\ell k}(x))$ and $\sum_{k=1}^K\lambda_k\Phi^{-1}(\psi_{\ell k}(x))$ are based on $\wh\bz$ and $\bz^*$, we denote them as $\sum_{k=1}^K\wh\lambda_k\Phi^{-1}(\widehat{\psi}_{\ell k}(x|\wh\bz))$ and $\sum_{k=1}^K\lambda_k^*\Phi^{-1}(\psi_{\ell k}(x|\bz^*))$ respectively.
In addition, we propose the oracle counterpart $\bar{\bg}=\sigma_{\ell}^*\sum_{k=1}^K\lambda_k^*\allowbreak\Phi^{-1}(\psi_{\ell k}(x|\wh\bz))$ of $\widehat{\bg}$.
Conditional on the event $\mathcal X_n$, all observations lie in $\mathcal{D}_{n,p}$, so the restricted bounds apply at every sample point.

\begin{enumerate}
\item[{\bf (1)}]

\noindent{\bf Step 1.} 
We first consider arbitrary $\bz\in\{1,\dots,K\}^n$. 
Note $\sum_{i=1}^n\mathbb{I}(X_{il}<x,z_{i}=k)=\sum_{i=1}^n\mathbb{I}(X_{il}<x)\mathbb{I}(z_{i}=k)$. 
Denote 
$\mathcal{Z}_k=\{\mathbb{I}\{w_{\btheta^*,k}(\bx)>w_{\btheta^*,j}(\bx)\;\forall j\ne k\}:\bx\in\mathcal{X}_n\},$ 
$\mathcal{I}=\{\mathbb{I}(X_1<x_1)\}\times\cdots\times\{\mathbb{I}(X_p<x_p)\},$
and $\mathcal{Z}=\mathcal{Z}_1\times\cdots\times\mathcal{Z}_K$. Since $\bg\in\mathcal{G}$ is monotone, the support of $\mathbb{I}\{w_{\btheta^*,k}(\bx)>w_{\btheta^*,j}(\bx),\;\forall j\ne k\}$ is a product of intervals in $\mathbb{R}$ (here we allow the intervals to include the negative and positive infinity), so $\text{VC}(\mathcal{Z}_k)<6$ and consequently $\text{VC}(\mathcal{Z})<6K$. Also, $\text{VC}(\mathcal{I})<3p$. 
Recall that for each $i=1,\dots,n$, conditional on $z_i^*$, $\bg^*(\bX_i)\sim\mathcal{N}(\bmu_{z_i^*}^*,\bSigma^*)$ and $g_{\ell}^*(X_{i\ell})\sim\mathcal{N}(\mu_{z_i^*\ell}^*,(\sigma_{\ell}^*)^2)$ for each $\ell=1,\dots,p$.  Then, since for each $\ell=1,\dots,p$ and some $\delta>0$, $\max_{\ell}\sum_{k=1}^K\lambda_k^*\int_{-\infty}^{\infty}x^{4+\delta}\phi(\{g_{\ell}^*(x)-\mu_k^*\}/\sigma_{\ell}^*)dg_{\ell}^*(x)<\infty$, we have $\max_{i,\ell}\mathbb{E}[|X_{i\ell}|^{4+\delta}]<\infty$. 
Then, for each $i=1,\dots,n$ and $\ell,\ell'=1,\dots,p$, we have
\begin{align*}
\mathbb{E}[|n^{-1}\sum_{i=1}^nX_{i\ell}X_{i\ell'}-n^{-1}\mathbb{E}[\sum_{i=1}^nX_{i\ell}X_{i\ell'}]|^2]
\le n^{-1}\mathbb{E}[X_{i\ell}^2X_{i\ell'}^2]
\le n^{-1}\max_{i,\ell}\mathbb{E}[X_{i\ell}^4].
\end{align*}
Then, for each $\ell=1,\dots,p$ and any $\zeta>0$, we have
\begin{align}
&~\mathbb{P}(|n^{-1}\sum_{i=1}^nX_{i\ell}X_{i\ell'}-n^{-1}\mathbb{E}[\sum_{i=1}^nX_{i\ell}X_{i\ell'}]|>\zeta)
\le\frac{\max_{i,\ell}\mathbb{E}[X_{i\ell}^4]}{n\zeta^2}.\label{eq:x_concentrate}
\end{align}
Then, by \eqref{eq:x_concentrate}, Theorem 9.3 in \citet{kosorok2008introduction}, and Corollary A.1 in \citet{chernozhukov2014gaussian},
\begin{align*}
\sup_PN(\mathcal{Z}_k,L_2(P),\varepsilon)\le\left(\frac{A}{\varepsilon}\right)^{6\eta_0K}\mbox{ and }
\sup_PN(\mathcal{I}\mathcal{Z}_k,L_2(P),\sqrt2\varepsilon)\le\left(\frac{A}{\varepsilon}\right)^{\eta_0(6K+3p)}.
\end{align*}
Hence, by Theorem 1.3 in \citet{talagrand1994sharper}, with probability at least $1-\zeta/24$ for any $\zeta>0$,
$\max_{1\le k\le K}|n^{-1}\sum_{i=1}^n\mathbb{I}(z_{i}=k)-\mathbb{P}(z_i=k)|\le L_nn^{-1/2}$ and
$\max_{1\le \ell\le p}\max_{1\le k\le K}\allowbreak\sup_{\bx\in\mathcal{D}_{n,p}}|n^{-1}\sum_{i=1}^n\mathbb{I}(X_{i\ell}\le x)\mathbb{I}(z_{i}=k)-\mathbb{P}(X_{\ell}\le x,z_i=k)|\le L_dn^{-1/2}$,
where $L_n=\widetilde{\eta}\{\log(nK/\zeta)\}^{1/2}$ and $L_d=\widetilde{\eta}[\log\{n(p+K)/\zeta\}]^{1/2}$. Thus, with probability at least $1-\zeta/12$ for any $\zeta>0$, for each $\bz\in\{1,\dots,K\}^n$,
\begin{align}
\max_{1\le \ell\le p}\max_{1\le k\le K}\sup_{\bx\in\mathcal{D}_{n,p}}\left|\frac{\sum_{i=1}^n\mathbb{I}(X_{i\ell}<x,z_{i}=k)}{\sum_{i=1}^n\mathbb{I}(z_{i}=k)}-\mathbb{P}\{X_{\ell}\le x|z_i=k\}\right|\le L_gn^{-1/2},\label{eqn:converg}
\end{align}
where $L_g=\min_k(\lambda_k^*)^{-1}(L_n+L_d)$. 

\noindent{\bf Step 2.} By Lemma \ref{lem:tail}, for  any ${\bz}\in\{1,\dots,K\}^n$ and $\bx\in\mathcal{X}_n$, with probability $1-\zeta$,
$$a_{n,p}\le\Phi(\{g_\ell^*(X_{i\ell})-\mu_{k\ell}^*\}/\sigma_\ell^*)\le1-a_{n,p}$$
for each $k=1,\dots,K$ and $\ell=1,\dots,p$. Taking $\Phi^{-1}$ at both sides, we have
$$-(L_4+|\mu_{k\ell}^*|)/\sigma_{\ell}^*\le\Phi^{-1}(\mathbb{P}(X_{1\ell}\le x|z_i=k))\le(L_4+|\mu_{k\ell}^*|)/\sigma_{\ell}^* $$
with probability $1-\zeta$, for each $k=1,\dots,K$ and $\ell=1,\dots,p$, where $L_4=\Phi^{-1}(1-a_{n,p})$.
In addition, by \eqref{eqn:converg}, with probability at least $1-7\zeta/12$,
\begin{align*}
&\;-L_gn^{-1/2}+\Phi(-\{\min_{\ell}\sigma_{\ell}^*\}^{-1}(L_4+\max_{k,\ell}|\mu_{k\ell}^*|))\\
\le&\; \max_{1\le \ell\le p}\max_{1\le k\le K}\{\sum_{i=1}^n\mathbb{I}(z_i=k)\}^
{-1}\sum_{i=1}^n\mathbb{I}(X_{i\ell}<x,z_i=k)\\
\le&\; L_gn^{-1/2}+\Phi(\{\min_{\ell}\sigma_{\ell}^*\}^{-1}(L_4+\max_{k,\ell}|\mu_{k\ell}^*|)). 
\end{align*}
Then, 
with probability at least $1-7\zeta/12$ for any $\zeta>0$,
\begin{align}
&~ \max_{1\le \ell\le p}\sup_{\bx\in\mathcal{D}_{n,p}}\Big|\Phi^{-1}(\wh\psi_{\ell k}(x)|\bz)\ -\Phi^{-1}(\psi_{\ell k}(x)|\bz)\Big|
\le\varepsilon'_g(n,p,s,\zeta),\label{eqn:converg3}
\end{align} where $\varepsilon'_g(n,p,s,\zeta)=\eta_{\delta,q}\min_k(\lambda_k^*)^{-1}(L_n+L_d)n^{-1/2}=\eta_{\delta,q}\min_k(\lambda_k^*)^{-1}(\widetilde{\eta}\{\log(nK/\delta)\}^{1/2}\allowbreak+\widetilde{\eta}[\log\{n(p+K)/\delta\}]^{1/2})n^{-1/2}$.
Thus, we have
\begin{align*}
\mathbb{P}\{\Vert\widehat{\bg}^{(0)}-\bar{\bg}^{(0)}\Vert_{\infty,s}\le \varepsilon_{\bg}(n,p,s,\zeta)\}> 1-\zeta/2.
\end{align*}

\item[{\bf (2)}]
With similar arguments in (1), we only consider the difference between  $\sum_{k=1}^K\lambda_k\Phi^{-1}(\allowbreak \psi_{\ell k}(x|\wh\bz))$ and $\sum_{k=1}^K\lambda_k^*\Phi^{-1}(\psi_{\ell k}(x|\bz^*))$ in remaining proof.

For each $\bz\in\{1,\dots,K\}^n$, since the indicator functions take value at $\{0,1\}$,
\begin{align*}
&\;\max_{1\le k\le K}\min_{\Pi\in\mathcal{S}_K}\left|\frac1n\sum_{i=1}^n\mathbb{I}(\Pi ({z}_{i})=k)
-\frac1n\sum_{i=1}^n\mathbb{I}(z_i^*=k)\right|\\
=&\;\max_{1\le k\le K}\min_{\Pi\in\mathcal{S}_K}\left|\frac1n\sum_{i=1}^n\{\mathbb{I}(\Pi ({z}_{i})=k,z_i^*\ne k)-\mathbb{I}(\Pi ({z}_{i})\ne k,z_i^*= k)\}\right|\\
\le&\;\max_{1\le k\le K}\min_{\Pi\in\mathcal{S}_K}\left|\frac1n\sum_{i=1}^n\{\mathbb{I}(\Pi ({z}_{i})=k,z_i^*\ne k)+\mathbb{I}(\Pi ({z}_{i})\ne k,z_i^*= k)\}\right|\\
\le&\;\min_{\Pi\in\mathcal{S}_K}\left\vert\frac1n\sum_{i=1}^n\mathbb{I}(\Pi ({z}_{i})\ne z_i^*)\right\vert.
\end{align*}
Also, 
\begin{align*}
&\;\max_{1\le \ell\le p}\max_{1\le k\le K}\sup_{\bx\in\mathcal{D}_{n,p}}\min_{\Pi\in\mathcal{S}_K}\left|\frac1n\sum_{i=1}^n\mathbb{I}(X_{i\ell}<x_{\ell},\Pi ({z}_{i})=k)
-\frac1n\sum_{i=1}^n\mathbb{I}(X_{i\ell}<x_{\ell},z_i^*=k)\right|\\
=&\;\max_{1\le \ell\le p}\max_{1\le k\le K}\sup_{\bx\in\mathcal{D}_{n,p}}\min_{\Pi\in\mathcal{S}_K}\bigg|\frac1n\sum_{i=1}^n\mathbb{I}(X_{i\ell}<x_{\ell})\{\mathbb{I}(\Pi ({z}_{i})=k,z_i^*\ne k)\\
&\;-\mathbb{I}(\Pi ({z}_{i})\ne k,z_i^*= k)\}\bigg|\\
=&\;\max_{1\le \ell\le p}\max_{1\le k\le K}\sup_{\bx\in\mathcal{D}_{n,p}}\min_{\Pi\in\mathcal{S}_K}\bigg|\frac1n\sum_{i=1}^n\{\mathbb{I}(X_{i\ell}<x_{\ell},\Pi ({z}_{i})=k,z_i^*\ne k)\\
&\;-\mathbb{I}(X_{i\ell}<x_{\ell},\Pi ({z}_{i})\ne k,z_i^*= k)\}\bigg|\\
\le&\;\max_{1\le \ell\le p}\max_{1\le k\le K}\sup_{\bx\in\mathcal{D}_{n,p}}\min_{\Pi\in\mathcal{S}_K}\bigg|\frac1n\sum_{i=1}^n\{\mathbb{I}(X_{i\ell}<x_{\ell},\Pi ({z}_{i})=k,z_i^*\ne k)\\
&\;+\mathbb{I}(X_{i\ell}<x_{\ell},\Pi ({z}_{i})\ne k,z_i^*= k)\}\bigg|\\
\le&\;\max_{1\le \ell\le p}\sup_{\bx\in\mathcal{D}_{n,p}}\min_{\Pi\in\mathcal{S}_K}\left\vert\frac1n\sum_{i=1}^n\mathbb{I}(X_{i\ell}<x_{\ell})\mathbb{I}(\Pi ({z}_{i})\ne z_i^*)\right\vert\\
\le&\;\min_{\Pi\in\mathcal{S}_K}\left\vert\frac1n\sum_{i=1}^n\mathbb{I}(\Pi ({z}_{i})\ne z_i^*)\right\vert.
\end{align*}
Since for each $\Pi \in\mathcal{S}_K$,
$\vert\frac1n\sum_{i=1}^n\mathbb{I}(\Pi ({z}_{i})\ne z_i^*)-\mathbb{E}[\mathbb{I}\{\Pi ({z}_{i})\ne z_i^*\}]\vert\le 4n^{-1/2}$ and $\vert\frac1n\sum_{i=1}^n\mathbb{I}(z_i^*=k)-\mathbb{E}[\mathbb{I}\{z_i^*=k\}]\vert\le 4n^{-1/2}$,
\begin{align*}
&\;\max_{1\le \ell\le p}\max_{1\le k\le K}\sup_{\bx\in\mathcal{D}_{n,p}}\min_{\Pi\in\mathcal{S}_K}\left|\frac{\sum_{i=1}^n\mathbb{I}(X_{i\ell}<x,\Pi ({z}_{i})=k)}{\sum_{i=1}^n\mathbb{I}(\Pi ({z}_{i})=k)}
-\frac{\sum_{i=1}^n\mathbb{I}(X_{i\ell}<x,z_i^*=k)}{\sum_{i=1}^n\mathbb{I}(z_i^*=k)}\right|\\
=&\;\max_{1\le \ell\le p}\max_{1\le k\le K}\sup_{\bx\in\mathcal{D}_{n,p}}\min_{\Pi\in\mathcal{S}_K}\Big|\frac{\sum_{i=1}^n\mathbb{I}(X_{i\ell}<x,\Pi ({z}_{i})=k)\sum_{i=1}^n\mathbb{I}(z_i^*=k)}{\sum_{i=1}^n\mathbb{I}(\Pi ({z}_{i})=k)\sum_{i=1}^n\mathbb{I}(z_i^*=k)}\\
&\qquad \qquad \qquad\qquad \qquad -\frac{\sum_{i=1}^n\mathbb{I}(X_{i\ell}<x,z_i^*=k)\sum_{i=1}^n\mathbb{I}(\Pi ({z}_{i})=k)}{\sum_{i=1}^n\mathbb{I}(\Pi ({z}_{i})=k)\sum_{i=1}^n\mathbb{I}(z_i^*=k)}\Big|\\
\le&\;\max_{1\le \ell\le p}\max_{1\le k\le K}\sup_{\bx\in\mathcal{D}_{n,p}}\min_{\Pi\in\mathcal{S}_K}\bigg\{\frac{\sum_{i=1}^n\mathbb{I}(z_i^*=k)|\sum_{i=1}^n\mathbb{I}(X_{i\ell}<x,\Pi ({z}_{i})=k)}{\sum_{i=1}^n\mathbb{I}(\Pi ({z}_{i})=k)\sum_{i=1}^n\mathbb{I}(z_i^*=k)}\\
&\;-\frac{\sum_{i=1}^n\mathbb{I}(X_{i\ell}<x,z_i^*=k)|}{\sum_{i=1}^n\mathbb{I}(\Pi ({z}_{i})=k)\sum_{i=1}^n\mathbb{I}(z_i^*=k)}\\
&\;+\frac{\sum_{i=1}^n\mathbb{I}(X_{i\ell}<x,z_i^*=k)|\sum_{i=1}^n\mathbb{I}(\Pi ({z}_{i})=k)-\sum_{i=1}^n\mathbb{I}(z_i^*=k)|}{\sum_{i=1}^n\mathbb{I}(\Pi ({z}_{i})=k)\sum_{i=1}^n\mathbb{I}(z_i^*=k)}\bigg\}\\
\le&\;\max_{1\le \ell\le p}\max_{1\le k\le K}\sup_{\bx\in\mathcal{D}_{n,p}}\min_{\Pi\in\mathcal{S}_K}\bigg\{\frac{\sum_{i=1}^n\mathbb{I}(z_i^*=k)|\sum_{i=1}^n\mathbb{I}(X_{i\ell}<x,\Pi ({z}_{i})=k)}{\max_{1\le k\le K}\sum_{i=1}^n\mathbb{I}(\Pi ({z}_{i})=k)\sum_{i=1}^n\mathbb{I}(z_i^*=k)}\\
&\;-\frac{\sum_{i=1}^n\mathbb{I}(X_{i\ell}<x,z_i^*=k)|}{\max_{1\le k\le K}\sum_{i=1}^n\mathbb{I}(\Pi ({z}_{i})=k)\sum_{i=1}^n\mathbb{I}(z_i^*=k)}\\
&\;+\frac{\sum_{i=1}^n\mathbb{I}(X_{i\ell}<x,z_i^*=k)|\sum_{i=1}^n\mathbb{I}(\Pi ({z}_{i})=k)-\sum_{i=1}^n\mathbb{I}(z_i^*=k)|}{\max_{1\le k\le K}\sum_{i=1}^n\mathbb{I}(\Pi ({z}_{i})=k)\sum_{i=1}^n\mathbb{I}(z_i^*=k)}\bigg\}\\
\le&\;\max_{1\le k\le K}\min_{\Pi\in\mathcal{S}_K}\frac{2|\frac1n\sum_{i=1}^n\mathbb{I}(\Pi ({z}_{i})\ne z_i^*)|}{\frac1{n^2}\max_{1\le k\le K}\sum_{i=1}^n\mathbb{I}(\Pi ({z}_{i})=k)\sum_{i=1}^n\mathbb{I}(z_i^*=k)}\\
\le&\;\frac{2\{\mathcal{L}({\bz})-\mathcal{L}(\widehat{\bz}_{\text{opt}})\}+n^{-1/2}+n^{-1}}{(\min_k\lambda_k^*)^2}\\
\le&\;\frac{3\{\mathcal{L}({\bz})-\mathcal{L}(\widehat{\bz}_{\text{opt}})\}}{(\min_k\lambda_k^*)^2}.
\end{align*}
Thus, 
{\small \begin{align}    
&~\max_{1\le \ell\le p}\max_{1\le k\le K}\sup_{\bx\in\mathcal{D}_{n,p}}\min_{\Pi\in\mathcal{S}_K}\left|\frac{\sum_{i=1}^n\mathbb{I}(X_{i\ell}<x,\Pi ({z}_{i})=k)}{\sum_{i=1}^n\mathbb{I}(\Pi ({z}_{i})=k)}-\frac{\sum_{i=1}^n\mathbb{I}(X_{i\ell}<x,z_i^*=k)}{\sum_{i=1}^n\mathbb{I}(z_i^*=k)}\right| \nonumber\\
\le&~ 3c_{\lambda}^{-2}\{\mathcal{L}({\bz})-\mathcal{L}(\widehat{\bz}_{\text{opt}})\}.\label{eqn:converg2}
\end{align}}

By \eqref{eqn:converg2}, 
\begin{align}
&\max_{1\le \ell\le p}\max_{1\le k\le K}\sup_{\bx\in\mathcal{D}_{n,p}}\min_{\Pi\in\mathcal{S}_K}\bigg|\Phi^{-1}(\frac{\sum_{i=1}^n\mathbb{I}(X_{i\ell}<x,\Pi ({z}_{i})=k)}{\sum_{i=1}^n\mathbb{I}(\Pi ({z}_{i})=k)})\nonumber\\
&\;-\Phi^{-1}(\frac{\sum_{i=1}^n\mathbb{I}(X_{i\ell}<x,z_i^*=k)}{\sum_{i=1}^n\mathbb{I}(z_i^*=k)})\bigg|\nonumber\\
\le& \max_{1\le \ell\le p}\max_{1\le k\le K}\sup_{\bx\in\mathcal{D}_{n,p}}\min_{\Pi\in\mathcal{S}_K}\{\phi(\Phi^{-1}(a_{n,p}))\}^{-1}\bigg|\frac{\sum_{i=1}^n\mathbb{I}(X_{i\ell}<x,\Pi ({z}_{i})=k)}{\sum_{i=1}^n\mathbb{I}(\Pi ({z}_{i})=k)}\nonumber\\
&\;-\frac{\sum_{i=1}^n\mathbb{I}(X_{i\ell}<x,z_i^*=k)}{\sum_{i=1}^n\mathbb{I}(z_i^*=k)}\bigg|\nonumber\\
\le& 3c_{\lambda}^{-2}\{\phi(\Phi^{-1}(a_{n,p}))\}^{-1}\{\mathcal{L}({\bz})-\mathcal{L}(\widehat{\bz}_{\text{opt}})\}.\label{eqn:converg2a}
\end{align}
The conclusion follows.

\end{enumerate}

\end{proof}

\bibliographystyle{apalike}
%bibliography{npn-cite} 

\end{document}